\documentstyle[12pt]{article}

\begin{document}
\onecolumn

\begin{titlepage}
\begin{center}
{\LARGE \bf Exact Solution for the Metric and the Motion \\
of Two Bodies in (1+1) Dimensional Gravity } \\ \vspace{2cm}
R.B. Mann\footnotemark\footnotetext{email: 
mann@avatar.uwaterloo.ca}\\
\vspace{1cm}
Dept. of Physics,
University of Waterloo
Waterloo, ONT N2L 3G1, Canada\\
and \\
\vspace{1cm}
T. Ohta \footnotemark\footnotetext{email:
t-oo1@ipc.miyakyo-u.ac.jp}\\
\vspace{1cm} 
Department of Physics, Miyagi University of Education,
Aoba-Aramaki, Sendai 980, Japan\\
\vspace{2cm}
\vspace{2cm}
\today\\
\end{center}
\begin{abstract}
We present the exact solution of two-body motion in (1+1) dimensional
dilaton gravity by solving the constraint equations in the canonical
formalism. The determining equation of the Hamiltonian is derived in
a transcendental form and the Hamiltonian is expressed for the system
of two identical particles in terms of the Lambert $W$ function. The $W$
function has two real branches which join smoothly onto each other
and the Hamiltonian on the principal branch reduces to the Newtonian
limit for small coupling constant. On the other branch the
Hamiltonian yields a new set of motions which can not be understood
as relativistically correcting the Newtonian motion. The explicit
trajectory in the phase space $(r, p)$ is illustrated for various
values of the energy. The analysis is extended to the case of 
unequal masses. The full
expression of metric tensor is given and the consistency between
the solution of the metric and the equations of motion is rigorously
proved.
\end{abstract}
\end{titlepage}
\onecolumn

\section{Introduction}

One of the oldest and most notoriously vexing problems
in gravitational theory is that of determining the (self-consistent) 
motion of $N$ bodies and the resultant metric they collectively
produce under their mutual gravitational influence \cite{EIH}.   
In general there is no exact solution to this problem
(although approximation techniques exist \cite{300yrs})
except in the case $N=2$ for Newtonian gravity, since 
energy dissipation in the form of gravitational
radiation obstructs attempts to obtain an exact $N=2$ solution.  
Only the static sector of the Hamiltonian has thus far been
determined exactly \cite{ookh}.

Lower dimensional theories of gravity do not contain gravitational
radiation and so offer the possibility of making useful progress
on this problem.  For example in $(2+1)$ dimensions,  the absence of 
a static gravitational  potential allows one to generalize 
the static 2-body metric to that of  two bodies moving with any 
speed \cite{Bellini}.  In $(1+1)$ dimensions one must necessarily
consider a dilatonic theory of gravity \cite{BanksMann}
since the Einstein tensor is topologically trivial in two dimensions.
One such theory in this class has been of particular interest insofar
as it has a consistent Newtonian limit \cite{r3}
(a problematic issue for a generic dilaton gravity theory \cite{jchan}),
allowing for the formulation of a general framework for deriving
a Hamiltonian for a system of particles \cite{OR}. In the slow 
motion, weak field limit this Hamiltonian coincides with that of
Newtonian gravity in $(1+1)$ dimensions. 

Motivated by the above, we consider here
the problem of the relativistic motion of 2 point masses under 
gravity in $(1+1)$ dimensions.  We work in the context of the dilatonic
theory of gravity mentioned above \cite{r3}. Both the classical
and quantum properties of this theory (referred to as $R=T$ theory)
have been extensively investigated \cite{r3,r4,r5,rtquant},
and it contains the Jackiw-Teitelboim lineal gravity theory \cite{JT}
as a special case \cite{r3}.
The specific form of the coupling of the dilaton field $\Psi$ to 
gravity is chosen so that it decouples from the classical field 
equations in such a way as to ensure that the evolution of the 
gravitational field is determined only 
by the matter stress-energy (and reciprocally) \cite{r3,r4}. It
thereby captures the essence of classical general relativity (as opposed to 
classical scalar-tensor theories) in two spacetime dimensions, and has 
$(1+1)$--dimensional analogs of many of its properties \cite{r4,r5}.  
Furthermore, this theory can be understood as the $D\rightarrow 2$ limit of 
general relativity (as opposed to some particular solution(s))
\cite{2dross}.

We consequently find that the motion of the bodies is governed entirely 
by their mutual gravitational influence, and that the spacetime metric 
is likewise fully determined by their stress-energy \cite{r3,r4}. 
Unlike the $(2+1)$ dimensional case, a Newtonian limit exists,
and there is a static gravitational potential.  The solution we
obtain gives the exact Hamiltonian to infinite order in the 
gravitational coupling constant. Hence the full structure of 
the theory from the weak field to the strong field limits can be
studied. While some of the phase-space trajectories we obtain can
be viewed as relativistic extensions of Newtonian motion, we find
that for sufficiently large values of the total energy a qualitatively
new set of trajectories arises that cannot be viewed in this way.

The outline of our paper is as follows. In section 2 we recapitulate
the derivation of the canonically reduced $N$-body Hamiltonian in
$(1+1)$ dimensions.  In sections 3 and 4 we solve the constraint equation
and then derive an expression for the exact Hamiltonian in the 2 body
case.  In section 5 we analyze the motion in the case of equal masses,
and in section 6 we consider the unequal mass case. In section 7 we
solve for the spacetime metric and in section 8 we investigate the
test-particle limit of our solution.  In section 9 we consider the
the dependence of the Newtonian limit on dimensionality. We
close our manuscript with some concluding remarks and directions
for further work.

\section{The canonically reduced $N$-body Hamiltonian}

First we shall review the outline of the canonical reduction of $(1+1)$ 
dimensional dilaton gravity \cite{OR}. The action integral for the 
gravitational field coupled with $N$ point particles is

\begin{eqnarray}\label{act1}
I&=&\int dx^{2}\left[
\frac{1}{2\kappa}\sqrt{-g}
\left\{\Psi R+\frac{1}{2}g^{\mu\nu}\nabla_{\mu}\Psi\nabla_{\nu}\Psi\right\}
\right.
\nonumber \\
&&\makebox[2em]{}-\left.\sum_{a=1}^{N} m_{a}\int d\tau_{a}
\left\{-g_{\mu\nu}(x)\frac{dz^{\mu}_{a}}{d\tau_{a}}
\frac{dz^{\nu}_{a}}{d\tau_{a}}\right\}^{1/2}\delta^{2}(x-z_{a}(\tau_{a}))
\right] 
\end{eqnarray}
where $\Psi$ is the dilaton field, $g_{\mu\nu}$ is the metric 
(with determinant $g$), $R$ is the Ricci scalar and $\tau_{a}$ is the proper 
time of $a$-th particle, with $\kappa=8\pi G/c^4$. The symbol 
$\nabla_{\mu}$ denotes the covariant derivative associated with 
$g_{\mu\nu}$.  

The field equations derived from the action (\ref{act1}) are
\begin{equation}\label{e2}
R-g^{\mu\nu}\nabla_{\mu}\nabla_{\nu}\Psi=0 
\end{equation}
\begin{equation}\label{e3}
\frac{1}{2}\nabla_{\mu}\Psi\nabla_{\nu}\Psi
-\frac{1}{4}g_{\mu\nu}\nabla^{\lambda}\Psi\nabla_{\lambda}\Psi
+g_{\mu\nu}\nabla^{\lambda}\nabla_{\lambda}\Psi
-\nabla_{\mu}\nabla_{\nu}\Psi=\kappa T_{\mu\nu} 
\end{equation}
\begin{equation}\label{geo}
\frac{d}{d\tau_{a}}
\left\{g_{\mu\nu}(z_{a})\frac{dz^{\nu}_{a}}{d\tau_{a}}\right\}
-\frac{1}{2}g_{\nu\lambda,\mu}(z_{a})\frac{dz^{\nu}_{a}}{d\tau_{a}}
\frac{dz^{\lambda}_{a}}{d\tau_{a}}=0 \;\;.
\end{equation}
where the stress-energy due to the point masses is
\begin{eqnarray}
T_{\mu\nu}=\sum_{a=1}^{N} m_{a}\int d\tau_{a}\frac{1}{\sqrt{-g}}
g_{\mu\sigma}g_{\nu\rho}\frac{dz^{\sigma}_{a}}{d\tau_{a}}
\frac{dz^{\rho}_{a}}{d\tau_{a}}\delta^{2}(x-z_{a}(\tau_{a}))\;\;.
\end{eqnarray}
Inserting the trace of (\ref{e3}) into (\ref{e2}) yields
%
%
%
\begin{equation}\label{RT}
R=\kappa T^{\mu}_{\;\;\mu} \;\;.
\end{equation}
Particle dynamics in $R=T$ theory may therefore be described in terms of the 
equations (\ref{geo}) and (\ref{RT}), which forms a closed system of 
equations for the gravity/matter system.

At first sight it may seem that 
the dynamics is independent of the dilaton field, since both (\ref{geo})
and (\ref{RT}) do not include $\Psi$. Note, however, that 
all three components of the metric tensor cannot be determined from 
(\ref{RT}), since it is only one equation. The two extra degrees of 
freedom are related to the choice of coordinates.
If the coordinate conditions are chosen to be independent of 
$\Psi$, equation (\ref{RT}) determines the metric tensor completely. 
However this need not be the case, and so more generally 
we need to know the dilaton 
field $\Psi$, through which the metric tensor is (indirectly) 
determined -- it
is this field that guarantees conservation of the stress-energy
tensor via (\ref{e3}). 

In the canonical formalism the action (\ref{act1}) is written in 
the form
\begin{equation}\label{act2}
I=\int dx^{2}\left\{\sum_{a}p_{a}\dot{z}_{a}\delta(x-z_{a}(x^{0}))
+\pi\dot{\gamma}+\Pi\dot{\Psi}+N_{0}R^{0}+N_{1}R^{1}\right\} 
\end{equation}
where  $\gamma=g_{11},  N_{0}= (-g^{00})^{-1/2}, N_{1}= g_{10}$,
$\pi$ and $\Pi$ are conjugate momenta to $\gamma$ and $\Psi$
respectively, and 
\begin{eqnarray}
R^{0}&=&-\kappa\sqrt{\gamma}\gamma\pi^{2}+2\kappa\sqrt{\gamma}\pi\Pi
+\frac{1}{4\kappa\sqrt{\gamma}}(\Psi^{\prime})^{2}
-\frac{1}{\kappa}\left(\frac{\Psi^{\prime}}{\sqrt{\gamma}}\right)^{\prime}
-\sum_{a}\sqrt{\frac{p^{2}_{a}}{\gamma}+m^{2}_{a}}\;
\delta(x-z_{a}(x^{0}))
\nonumber \\ \label{R0}
\\
R^{1}&=&\frac{\gamma^{\prime}}{\gamma}\pi-\frac{1}{\gamma}\Pi\Psi^{\prime}
+2\pi^{\prime}
+\sum_{a}\frac{p_{a}}{\gamma}\delta(x-z_{a}(x^{0})) \;\;.\label{R1}
\end{eqnarray}
with the symbols $(\;\dot{}\;)$ and  $(\;^{\prime}\;)$
denoting $\partial_{0}$ and $\partial_{1}$, respectively.

The transformation from (\ref{act1}) to (\ref{act2})
is carried out by using the decomposition of the scalar curvature
in terms of the extrinsic curvature $K$ via
$$
\sqrt{-g}R=-2\partial_{0}
(\sqrt{\gamma}K)+2\partial_{1}[\sqrt{\gamma}(N^{1}K-\gamma^{-1}\partial
_{1}N_{0})]
$$ 
where $K=(2N_{0}\gamma)^{-1}(2\partial_{1}N_{1}
-\gamma^{-1}N_{1}\partial_{1}\gamma -\partial_{0}\gamma)$ 
and changing the particle Lagrangian into first order form.

The action (\ref{act2}) leads to the system of field equations :
\begin{eqnarray}
\dot{\pi}&+&N_{0}\left\{\frac{3\kappa}{2}\sqrt{\gamma}\pi^{2}
-\frac{\kappa}{\sqrt{\gamma}}\pi\Pi
+\frac{1}{8\kappa\sqrt{\gamma}\gamma}(\Psi^{\prime})^{2}
-\sum_{a}\frac{p^{2}_{a}}{2\gamma^{2}\sqrt{\frac{p^{2}_{a}}{\gamma}
+m^{2}_{a}}}\;\delta(x-z_{a}(x^{0}))\right\}
\nonumber \\
&+&N_{1}\left\{-\frac{1}{\gamma^{2}}\Pi\Psi^{\prime}
+\frac{\pi^{\prime}}{\gamma}
+\sum_{a}\frac{p_{a}}{\gamma^{2}}\;\delta(x-z_{a}(x^{0}))\right\}
+N^{\prime}_{0}\frac{1}{2\kappa\sqrt{\gamma}\gamma}\Psi^{\prime}
+N^{\prime}_{1}\frac{\pi}{\gamma}=0
\nonumber \label{feq1} 
\\
\\ &&\makebox[5em]{}\dot{\gamma}-N_{0}(2\kappa\sqrt{\gamma}\gamma\pi
-2\kappa\sqrt{\gamma}\Pi)
+N_{1}\frac{\gamma^{\prime}}{\gamma}-2N^{\prime}_{1}=0 \label{feq2} 
\\
&&\makebox[5em]{}R^{0}=0 \label{feq3} 
\\
&&\makebox[5em]{}R^{1}=0 \label{feq4}
\\
&&\makebox[5em]{}\dot{\Pi}+\partial_{1}(-\frac{1}{\gamma}N_{1}\Pi
+\frac{1}{2\kappa\sqrt{\gamma}}N_{0}\Psi^{\prime}
+\frac{1}{\kappa\sqrt{\gamma}}N^{\prime}_{0})=0 \label{feq5}
\\
&&\makebox[5em]{}
\dot{\Psi}+N_{0}(2\kappa\sqrt{\gamma}\pi)-N_{1}(\frac{1}{\gamma}
\Psi^{\prime})=0 \label{feq6}
\\
&&\dot{p}_{a}+\frac{\partial N_{0}}{\partial z_{a}}\sqrt{\frac{p^{2}_{a}}
{\gamma}+m^{2}_{a}}-\frac{N_{0}}{2\sqrt{\frac{p^{2}_{a}}{\gamma}+m^{2}_{a}}}
\frac{p^{2}_{a}}{\gamma^{2}}\frac{\partial\gamma}{\partial z_{a}}
-\frac{\partial N_{1}}{\partial z_{a}}\frac{p_{a}}{\gamma}
+N_{1}\frac{p_{a}}{\gamma^{2}}\frac{\partial\gamma}{\partial z_{a}}=0 
\label{feq7}
\\
&&\makebox[5em]{}\dot{z_{a}}-N_{0}\frac{\frac{p_{a}}{\gamma}}
{\sqrt{\frac{p^{2}_{a}}{\gamma}+m^{2}_{a}}}
+\frac{N_{1}}{\gamma}=0 \;\;.\label{feq8}
\end{eqnarray}
In the equations (\ref{feq7}) and (\ref{feq8}), all metric components 
($N_{0}$, $N_{1}$, $\gamma$) are evaluated at the point $x=z_{a}$ and
\[\frac{\partial f}{\partial z_{a}}\equiv 
\left.\frac{\partial f(x)}{\partial x}\right|_{x=z_{a}}
\]
This system of equations is equivalent to the set of equations
(\ref{e2}), (\ref{e3}) and (\ref{geo}). 

The action (\ref{act2}) also shows that $N_{0}$ and $N_{1}$ are 
Lagrange multipliers, and the equations (\ref{feq3}) and (\ref{feq4}) 
are constraints. We may investigate the canonical structure of the theory 
via the generator which arises from the variation 
of the action at the boundaries :
\begin{equation}\label{gen1}
G=\int dx\left\{\sum_{a}p_{a}\delta(x-z_{a})\delta z_{a}
+\pi\delta h-\Psi\delta\Pi\right\} \;\;\;,
\end{equation}
where $h\equiv 1+\gamma$. This form was obtained by adding a total
time derivative $-\partial_{0}(\Pi\Psi)$ to the original action
(\ref{act2}) and taking the constraints into account. 
Since the only linear terms in the constraints are 
$(\Psi^{\prime}/\sqrt{\gamma})^{\prime}$ and $\pi^{\prime}$, we may 
solve for these quantities in terms of the dynamical and gauge 
({\it i.e.} co-ordinate) degrees of freedom. Bearing this fact in 
mind, we transform the generator (\ref{gen1}) to
\begin{eqnarray}\label{gen2}
G&=&\int dx\left\{\sum_{a}p_{a}\delta(x-z_{a})\delta z_{a}
-\left[-\frac{1}{\kappa}\left(\frac{\Psi^{\prime}}{\sqrt{h-1}}
\right)^{\prime}\right]
\delta\left[-\frac{\kappa}{\triangle}\left(\sqrt{h-1}\frac{1}{\triangle}
\Pi^{\prime}\right)^{\prime}\right]
\right.
\nonumber \\
&&\makebox[5em]{}\left.-\left[2\pi^{\prime}-\left(\frac{\Psi^{\prime}}{h-1}
\right)^{\prime}\frac{1}{\triangle}\Pi^{\prime}-\frac{1}{h-1}\Pi\Psi^{\prime}
\right]\delta\left(\frac{1}{2\triangle}h^{\prime}\right)\right\} \;\;.
\label{e22}
\end{eqnarray}
where $1/\triangle$ is the inverse of the operator 
$\triangle=\partial^{2}/\partial x^{2}$ 
with appropriate boundary condition, and we have discarded surface 
terms.
 
Adopting the coordinate conditions
\begin{equation}\label{cc1}
x=\frac{1}{2\triangle}h^{\prime} 
\qquad
t=-\frac{\kappa}{\triangle}\left(\sqrt{h-1}\frac{1}{\triangle}\Pi^{\prime}
\right)^{\prime} \;\;
\end{equation}
allows the generator (\ref{gen2}) to be expressed in the canonical form
\begin{equation}
G=\int dx\left\{\sum_{a}p_{a}\delta(x-z_{a})\delta z_{a}
-\cal T\mit_{0 \mu}\delta x^{\mu}\right\} 
\end{equation}
where
\begin{eqnarray}
\cal T\mit_{0 0}&=& \cal H\mit =-\frac{1}{\kappa}\triangle\rm\Psi\mit
\\
\cal T\mit_{0 1}&=&2\pi^{\prime} \;\;.
\end{eqnarray}
In deriving this expression for $\cal T\mit_{0 \mu}$, we made use of
the differential forms of the coordinate conditions
\begin{equation}\label{cc2}
\gamma=1 \qquad \Pi=0 \;\;. 
\end{equation}
As discussed in \cite{OR}, consistency between the integral 
form (\ref{cc1}) of the coordinate conditions and the differential
form (\ref{cc2}) is assured when one retains the appropriate 
boundary conditions for the integral operator $1/\triangle$ and 
takes a limiting procedure by introducing a regulator.
Thus, $\cal H\mit$ is the Hamiltonian density of the system 
and $\cal T\mit_{0 1}$ is the momentum density. 

The action (\ref{act2}) is similarly tranformed and reduces to
\begin{equation}
I=\int dx^{2}\left\{\sum_{a}p_{a}\dot{z}_{a}\delta(x-z_{a})
-\cal H\mit\right\}\;\;.
\end{equation}

\noindent
Thus the reduced Hamiltonian for the system of particles is  
\begin{equation}\label{ham1}
H=\int dx \cal H\mit =-\frac{1}{\kappa}\int dx \triangle\Psi 
\end{equation}
where $\Psi$ is a function of $z_{a}$ and $p_{a}$ and is determined
by solving the constraints which are under the coordinate conditions
(\ref{cc2})
\begin{equation}\label{cst1}
\triangle\Psi-\frac{1}{4}(\Psi^{\prime})^{2}
+\kappa^{2}\pi^{2}+\kappa\sum_{a}\sqrt{p^{2}_{a}+m^{2}_{a}}
\delta(x-z_{a})=0
\end{equation}
\begin{equation}\label{cst2}
2\pi^{\prime}+\sum_{a}p_{a}\delta(x-z_{a})=0 \;\;.
\end{equation}
The expression of the Hamiltonian (\ref{ham1}) is analogous to the 
reduced Hamiltonian in $(3+1)$ dimensional general relativity.
The proof of the consistency of this canonical reduction was given
in \cite{OR}: namely the canonical equations of motion derived 
from the reduced Hamiltonian (\ref{ham1}) are identical with the 
equations (\ref{feq7}) and (\ref{feq8}).

\section{Matching conditions and the solution to the constraint
equations}

The standard approach for investigating the dynamics of a system of 
particles is to derive an explicit expression for the Hamiltonian,
in which all information on the motion of the particles is
included. In this section we solve the constraints (\ref{cst1}) and 
(\ref{cst2}) for the system of two particles and determine the 
Hamiltonian (\ref{cst2}).

Defining $\phi$ and $\chi$ by
\begin{equation}
\Psi=-4\mbox{log}|\phi| \qquad \pi=\chi^{\prime}
\end{equation}
the constraints (\ref{cst1}) and (\ref{cst2}) for a two-particle system 
become
\begin{eqnarray}
\triangle\phi-\frac{\kappa^{2}}{4}(\chi^{\prime})^{2}\phi
&=&\frac{\kappa}{4}
\left\{\sqrt{p^{2}_{1}+m^{2}_{1}}\;\phi(z_{1})\delta(x-z_{1})
+\sqrt{p^{2}_{2}+m^{2}_{2}}\;\phi(z_{2})\delta(x-z_{2})\right\}
\label{phi-eq}
\\
\triangle\chi &=&-\frac{1}{2}\left\{p_{1}\delta(x-z_{1})
+p_{2}\delta(x-z_{2})\right\}\;\;. \label{chi-eq}
\end{eqnarray}
The general solution to (\ref{chi-eq}) is
\begin{equation}
\chi=-\frac{1}{4}\left\{p_{1}\mid x-z_{1}\mid+p_{2}\mid x-z_{2}\mid\right\}
-\epsilon Xx+\epsilon C_{\chi} \;\;.
\end{equation}
The factor $\epsilon$ ($\epsilon^{2}=1$) has been introduced in
the constants $X$ and $C_{\chi}$ so that the T-inversion 
(time reversal) properties of $\chi$ are explicitly manifest. 
By definition, $\epsilon$ changes sign under time reversal
and so, therefore, does $\chi$.
 
Consider first the case  $z_{2}<z_{1}$, for which we may divide 
spacetime into three regions:
\[
z_{1}<x \makebox[3em]{}\mbox{(+) region}
\]
\[
z_{2}<x<z_{1} \makebox[1.5em]{}\mbox{(0) region}
\]
\[
x<z_{2} \makebox[3em]{}\mbox{(-) region}
\]
In each region $\chi^{\prime}$ is constant:
\begin{equation}
\chi^{\prime}=\left\{
\begin{array}{ll}
-\epsilon X-\frac{1}{4}(p_{1}+p_{2}) & \makebox[3em]{}\mbox{(+) region} \\
-\epsilon X+\frac{1}{4}(p_{1}-p_{2}) & \makebox[3em]{}\mbox{(0) region} \\
-\epsilon X+\frac{1}{4}(p_{1}+p_{2}) & \makebox[3em]{}\mbox{(-) region}
\end{array}
\right.
\end{equation}
General solutions to the homogeneous equation 
$\triangle\phi-(\kappa^{2}/4)(\chi^{\prime})^{2}\phi=0$ in each
region are 
\begin{equation}
\left\{
\begin{array}{l}
\phi_{+}(x)=A_{+}e^{\frac{\kappa}{2}\left\{X+\frac{\epsilon}{4}
(p_{1}+p_{2})\right\}x}
+B_{+}e^{-\frac{\kappa}{2}\left\{X+\frac{\epsilon}{4}
(p_{1}+p_{2})\right\}x} \\
\phi_{0}(x)=A_{0}e^{\frac{\kappa}{2}\left\{X-\frac{\epsilon}{4}
(p_{1}-p_{2})\right\}x}
+B_{0}e^{-\frac{\kappa}{2}\left\{X-\frac{\epsilon}{4}
(p_{1}-p_{2})\right\}x} \\
\phi_{-}(x)=A_{-}e^{\frac{\kappa}{2}\left\{X-\frac{\epsilon}{4}
(p_{1}+p_{2})\right\}x}
+B_{-}e^{-\frac{\kappa}{2}\left\{X-\frac{\epsilon}{4}
(p_{1}+p_{2})\right\}x}
\end{array}
\right. \;\;.
\end{equation}
For these solutions to be the actual solutions to Eq.(\ref{phi-eq}) 
with delta function source terms, they must satisfy the following
matching conditions at $x=z_{1}, z_{2}$:
{
\setcounter{enumi}{\value{equation}}
\addtocounter{enumi}{1}
\setcounter{equation}{0}
\renewcommand{\theequation}{\theenumi\alph{equation}}
\begin{eqnarray}
&&\phi_{+}(z_{1})=\phi_{0}(z_{1})=\phi(z_{1}) \label{match1}\\
&&\phi_{-}(z_{2})=\phi_{0}(z_{2})=\phi(z_{2}) \label{match2}\\
&&\phi^{\prime}_{+}(z_{1})-\phi^{\prime}_{0}(z_{1})
=\frac{\kappa}{4}\sqrt{p^{2}_{1}+m^{2}_{1}}\phi(z_{1}) \label{match3}\\
&&\phi^{\prime}_{0}(z_{2})-\phi^{\prime}_{-}(z_{2})
=\frac{\kappa}{4}\sqrt{p^{2}_{2}+m^{2}_{2}}\phi(z_{2}) \;\;.\label{match4}
\end{eqnarray}
\setcounter{equation}{\value{enumi}}
}
The conditions (\ref{match1}) and (\ref{match3}) lead to
\begin{eqnarray}
\lefteqn{
e^{\frac{\kappa}{2}\left\{X+\frac{\epsilon}{4}(p_{1}+p_{2})
\right\}z_{1}}A_{+}
+e^{-\frac{\kappa}{2}\left\{X+\frac{\epsilon}{4}(p_{1}+p_{2})
\right\}z_{1}}B_{+}}
\nonumber \\
&=&e^{\frac{\kappa}{2}\left\{X-\frac{\epsilon}{4}
(p_{1}-p_{2})\right\}z_{1}}A_{0}
+e^{-\frac{\kappa}{2}\left\{X-\frac{\epsilon}{4}(p_{1}-p_{2})
\right\}z_{1}}B_{0}
\end{eqnarray}
and
\begin{eqnarray}
\lefteqn{
e^{\frac{\kappa}{2}\left\{X+\frac{\epsilon}{4}
(p_{1}+p_{2})\right\}z_{1}}A_{+}
-e^{-\frac{\kappa}{2}\left\{X+\frac{\epsilon}{4}
(p_{1}+p_{2})\right\}z_{1}}B_{+}}
\nonumber \\
&=&\frac{\sqrt{p^{2}_{1}+m^{2}_{1}}+2\left\{X-\frac{\epsilon}{4}
(p_{1}-p_{2})\right\}}
{2\left\{X+\frac{\epsilon}{4}(p_{1}+p_{2})\right\}}
e^{\frac{\kappa}{2}\left\{X-\frac{\epsilon}{4}
(p_{1}-p_{2})\right\}z_{1}}A_{0}
\nonumber \\
&&+\frac{\sqrt{p^{2}_{1}+m^{2}_{1}}-2\left\{X-\frac{\epsilon}{4}
(p_{1}-p_{2})\right\}}
{2\left\{X+\frac{\epsilon}{4}(p_{1}+p_{2})\right\}}
e^{-\frac{\kappa}{2}\left\{X-\frac{\epsilon}{4}
(p_{1}-p_{2})\right\}z_{1}}B_{0}
\end{eqnarray}
and then
{
\setcounter{enumi}{\value{equation}}
\addtocounter{enumi}{1}
\setcounter{equation}{0}
\renewcommand{\theequation}{\theenumi\alph{equation}}
\begin{eqnarray}
A_{+}&=&\frac{\sqrt{p^{2}_{1}+m^{2}_{1}}+4X+\epsilon p_{2}}
{4X+\epsilon(p_{1}+p_{2})}
e^{-\frac{\kappa\epsilon}{4}p_{1}z_{1}}A_{0}
+\frac{\sqrt{p^{2}_{1}+m^{2}_{1}}+\epsilon p_{1}}
{4X+\epsilon(p_{1}+p_{2})}
e^{-\kappa\left(X+\frac{\epsilon}{4}p_{2}\right)z_{1}}B_{0} \label{A+0}\\
B_{+}&=&-\frac{\sqrt{p^{2}_{1}+m^{2}_{1}}-\epsilon p_{1}}
{4X+\epsilon(p_{1}+p_{2})}
e^{\kappa\left(X+\frac{\epsilon}{4}p_{2}\right)z_{1}}A_{0}
-\frac{\sqrt{p^{2}_{1}+m^{2}_{1}}-4X-\epsilon p_{2}}
{4X+\epsilon(p_{1}+p_{2})}
e^{\frac{\kappa\epsilon}{4}p_{1}z_{1}}B_{0} \;\;.
\end{eqnarray}
\setcounter{equation}{\value{enumi}}
}
Similarly the conditions (\ref{match2}) and (\ref{match4}) lead to
\begin{eqnarray}
\lefteqn{
e^{\frac{\kappa}{2}\left\{X-\frac{\epsilon}{4}
(p_{1}+p_{2})\right\}z_{2}}A_{-}
+e^{-\frac{\kappa}{2}\left\{X-\frac{\epsilon}{4}
(p_{1}+p_{2})\right\}z_{2}}B_{-}}
\nonumber \\
&=&e^{\frac{\kappa}{2}\left\{X-\frac{\epsilon}{4}
(p_{1}-p_{2})\right\}z_{2}}A_{0}
+e^{-\frac{\kappa}{2}\left\{X-\frac{\epsilon}{4}
(p_{1}-p_{2})\right\}z_{2}}B_{0}
\end{eqnarray}
and
\begin{eqnarray}
\lefteqn{
-e^{\frac{\kappa}{2}\left\{X-\frac{\epsilon}{4}
(p_{1}+p_{2})\right\}z_{2}}A_{-}
+e^{-\frac{\kappa}{2}\left\{X-\frac{\epsilon}{4}
(p_{1}+p_{2})\right\}z_{2}}B_{-}}
\nonumber \\
&=&\frac{\sqrt{p^{2}_{2}+m^{2}_{2}}-2\left\{X-\frac{\epsilon}{4}
(p_{1}-p_{2})\right\}}
{2\left\{X-\frac{\epsilon}{4}(p_{1}+p_{2})\right\}}
e^{\frac{\kappa}{2}\left\{X-\frac{\epsilon}{4}(p_{1}-p_{2})\right\}
z_{2}}A_{0}
\nonumber \\
&&+\frac{\sqrt{p^{2}_{2}+m^{2}_{2}}+2\left\{X-\frac{\epsilon}{4}
(p_{1}-p_{2})\right\}}
{2\left\{X-\frac{\epsilon}{4}(p_{1}+p_{2})\right\}}
e^{-\frac{\kappa}{2}\left\{X-\frac{\epsilon}{4}(p_{1}-p_{2})\right\}
z_{2}}B_{0}
\end{eqnarray}
and then
{
\setcounter{enumi}{\value{equation}}
\addtocounter{enumi}{1}
\setcounter{equation}{0}
\renewcommand{\theequation}{\theenumi\alph{equation}}
\begin{eqnarray}
A_{-}&=&-\frac{\sqrt{p^{2}_{2}+m^{2}_{2}}-4X+\epsilon p_{1}}
{4X-\epsilon(p_{1}+p_{2})}
e^{\frac{\kappa\epsilon}{4}p_{2}z_{2}}A_{0}
-\frac{\sqrt{p^{2}_{2}+m^{2}_{2}}+\epsilon p_{2}}
{4X-\epsilon(p_{1}+p_{2})}
e^{-\kappa\left(X-\frac{\epsilon}{4}p_{1}\right)z_{2}}B_{0} \\
B_{-}&=&\frac{\sqrt{p^{2}_{2}+m^{2}_{2}}-\epsilon p_{2}}
{4X-\epsilon(p_{1}+p_{2})}
e^{\kappa\left(X-\frac{\epsilon}{4}p_{1}\right)z_{2}}A_{0}
+\frac{\sqrt{p^{2}_{2}+m^{2}_{2}}+4X-\epsilon p_{1}}
{4X-\epsilon(p_{1}+p_{2})}
e^{-\frac{\kappa\epsilon}{4}p_{2}z_{2}}B_{0} \;\;. \label{B-0}
\end{eqnarray}
\setcounter{equation}{\value{enumi}}
}

Since the magnitudes of both $\phi$ and $\chi$ increase with 
increasing $|x|$, it is necessary to impose a boundary condition
which guarantees that the surface terms which arise in transforming the 
action vanish and simultaneously preserves the finiteness
of the Hamiltonian. 

In the iterative analysis in \cite{OR} this condition has been shown
to be 
\begin{equation}
\Psi^{2}-4\kappa^{2}\chi^{2}=0 \makebox[2em]{}
\mbox{in the region}\makebox[1em]{}z_{1}<x \makebox[1em]{}\mbox{and}
\makebox[1em]{} x<z_{2} \;\;.
\end{equation}
Since
\begin{equation}
\chi=\left\{
\begin{array}{ll}
-\left\{\epsilon X+\frac{1}{4}(p_{1}+p_{2})\right\}x+\epsilon C_{\chi}
+\frac{1}{4}(p_{1}z_{1}+p_{2}z_{2}) & \makebox[3em]{}(+) \mbox{region} \\
-\left\{\epsilon X-\frac{1}{4}(p_{1}+p_{2})\right\}x+\epsilon C_{\chi}
-\frac{1}{4}(p_{1}z_{1}+p_{2}z_{2}) & \makebox[3em]{}(-) \mbox{region}
\end{array} \right.
\end{equation}
the boundary condition implies
\begin{eqnarray}
&&A_{-}=B_{+}=0 \label{bound1}
\\
&&-\mbox{log}A_{+}-\frac{\kappa\epsilon}{8}(p_{1}z_{1}+p_{2}z_{2})=
\mbox{log}B_{-}+\frac{\kappa\epsilon}{8}(p_{1}z_{1}+p_{2}z_{2})
=\frac{\kappa}{2}C_{\chi} \;\;. \label{bound2}
\end{eqnarray}
The condition (\ref{bound1}) leads to
\begin{equation}\label{AB01}
\frac{A_{0}}{B_{0}}=-\frac{\sqrt{p^{2}_{2}+m^{2}_{2}}+\epsilon p_{2}}
{\sqrt{p^{2}_{2}+m^{2}_{2}}-4X+\epsilon p_{1}}
e^{-\kappa\left\{X-\frac{\epsilon}{4}(p_{1}-p_{2})\right\}z_{2}} 
\end{equation}
and
\begin{equation}\label{AB02}
\frac{A_{0}}{B_{0}}=-\frac{\sqrt{p^{2}_{1}+m^{2}_{1}}-4X-\epsilon p_{2}}
{\sqrt{p^{2}_{1}+m^{2}_{1}}-\epsilon p_{1}}
e^{-\kappa\left\{X-\frac{\epsilon}{4}(p_{1}-p_{2})\right\}z_{1}} \;\;.
\end{equation}
{}From (\ref{AB01}) and (\ref{AB02}) we have
\begin{eqnarray}\label{eqX1}
&&(\sqrt{p^{2}_{1}+m^{2}_{1}}-\epsilon p_{2}-4X)
(\sqrt{p^{2}_{2}+m^{2}_{2}}+\epsilon p_{1}-4X)
\nonumber \\
&& \makebox[5em]{}=(\sqrt{p^{2}_{1}+m^{2}_{1}}-\epsilon p_{1})
(\sqrt{p^{2}_{2}+m^{2}_{2}}+\epsilon p_{2})
e^{\kappa\left\{X-\frac{\epsilon}{4}(p_{1}-p_{2})\right\}
(z_{1}-z_{2})} \;\;.
\end{eqnarray}
On the other hand the condition (\ref{bound2}) leads to
\begin{eqnarray}\label{AB03}
&&\left\{\frac{\sqrt{p^{2}_{1}+m^{2}_{1}}+4X+\epsilon p_{2}}
{4X+\epsilon(p_{1}+p_{2})}
e^{-\frac{\kappa\epsilon}{4}p_{1}z_{1}}A_{0}
+\frac{\sqrt{p^{2}_{1}+m^{2}_{1}}+\epsilon p_{1}}
{4X+\epsilon(p_{1}+p_{2})}
e^{-\kappa\left(X+\frac{\epsilon}{4}p_{2}\right)z_{1}}B_{0} \right\}
\nonumber \\
&& \times\left\{\frac{\sqrt{p^{2}_{2}+m^{2}_{2}}-\epsilon p_{2}}
{4X-\epsilon(p_{1}+p_{2})}
e^{\kappa\left(X-\frac{\epsilon}{4}p_{1}\right)z_{2}}A_{0}
+\frac{\sqrt{p^{2}_{2}+m^{2}_{2}}+4X-\epsilon p_{1}}
{4X-\epsilon(p_{1}+p_{2})}
e^{-\frac{\kappa\epsilon}{4}p_{2}z_{2}}B_{0}\right\}
=e^{-\frac{\kappa\epsilon}{4}(p_{1}z_{1}+p_{2}z_{2})} \;\;.
\nonumber \\
\end{eqnarray}

Using the notation
\begin{eqnarray}
M_{1}&\equiv& \sqrt{p^{2}_{1}+m^{2}_{1}}-\epsilon p_{1}
\nonumber \\
M_{2}&\equiv& \sqrt{p^{2}_{2}+m^{2}_{2}}+\epsilon p_{2}
\nonumber \\
L_{1}&\equiv& 4X-\epsilon p_{1}-\sqrt{p^{2}_{2}+m^{2}_{2}}
\nonumber \\
L_{2}&\equiv& 4X+\epsilon p_{2}-\sqrt{p^{2}_{1}+m^{2}_{1}}
\\
L_{+}&\equiv& 4X+\epsilon(p_{1}+p_{2})
\nonumber \\
L_{0}&\equiv& 4X-\epsilon p_{1}+\epsilon p_{2}
\nonumber \\
L_{-}&\equiv& 4X-\epsilon(p_{1}+p_{2})\;\;.
\nonumber 
\end{eqnarray}
we obtain %
\begin{eqnarray}
A_{0}&=&\frac{(M_{2}L_{2})^{1/2}}{L_{0}}
e^{-\frac{\kappa}{8}L_{0}z_{2}}
=\left(\frac{L_{1}}{M_{1}}\right)^{1/2}\frac{L_{2}}{L_{0}}
e^{-\frac{\kappa}{8}L_{0}z_{1}}\label{A0} 
\\
B_{0}&=&\left(\frac{L_{2}}{M_{2}}\right)^{1/2}\frac{L_{1}}{L_{0}}
e^{\frac{\kappa}{8}L_{0}z_{2}}
=\frac{(M_{1}L_{1})^{1/2}}{L_{0}}
e^{\frac{\kappa}{8}L_{0}z_{1}}
\label{B0} 
\end{eqnarray}
from (\ref{AB01}), (\ref{AB02}) and (\ref{AB03}). 
Substituting (\ref{A0}) and (\ref{B0}) into (\ref{A+0}) and (\ref{B-0}) 
we get
\begin{eqnarray}
A_{+}&=&\left(\frac{L_{1}}{M_{1}}\right)^{1/2}
e^{-\frac{\kappa}{8}L_{+}z_{1}}
\label{A+} \\
B_{-}&=&\left(\frac{L_{2}}{M_{2}}\right)^{1/2}
e^{\frac{\kappa}{8}L_{-}z_{2}}
\label{B-}
\end{eqnarray}
and then
\begin{eqnarray}
\phi_{+}&=&\left(\frac{L_{1}}{M_{1}}\right)^{1/2}
e^{\frac{\kappa}{8}L_{+}(x-z_{1})}
\label{phi+} \\
\phi_{-}&=&\left(\frac{L_{2}}{M_{2}}\right)^{1/2}
e^{-\frac{\kappa}{8}L_{-}(x-z_{2})} 
\label{phi-} \\
\phi_{0}&=&\frac{1}{L_{0}}\left(\frac{L_{1}L_{2}}{M_{1}M_{2}}\right)^{1/2}
\left\{(M_{2}L_{2})^{1/2}e^{\frac{\kappa}{8}L_{0}(x-z_{1})}
+(M_{1}L_{1})^{1/2}e^{-\frac{\kappa}{8}L_{0}(x-z_{2})}
\right\} \;\;.\label{phi0} 
\end{eqnarray}

In the case of $z_{1}<z_{2}$ we have to interchange the suffices 1 and 2.
The equation (\ref{eqX1}) which determines $X$ then generalizes to 
\begin{eqnarray}\label{eqX2}
&&(\sqrt{p^{2}_{1}+m^{2}_{1}}-\epsilon\tilde{p}_{2}-4X)
(\sqrt{p^{2}_{2}+m^{2}_{2}}+\epsilon\tilde{p}_{1}-4X)
\nonumber \\
&&\makebox[5em]{}=(\sqrt{p^{2}_{1}+m^{2}_{1}}-\epsilon\tilde{p}_{1})
(\sqrt{p^{2}_{2}+m^{2}_{2}}+\epsilon\tilde{p}_{2})
e^{\kappa\left\{X-\frac{\epsilon}{4}(\tilde{p}_{1}-\tilde{p}_{2})\right\}|r|}
\end{eqnarray}
where $r\equiv z_{1}-z_{2}$ and $\tilde{p}_{a}\equiv p_{a}\mbox{sgn}
(z_{1}-z_{2})$.

\section{Determining equation for the Hamiltonian and the canonical
equations of motion}

Since the solutions of $\phi$ give
\begin{equation}
\frac{\phi^{\prime}_{+}(x)}{\phi_{+}(x)}
=\frac{\kappa}{2}\left\{X+\frac{\epsilon}{4}(p_{1}+p_{2})\right\}
\makebox[2em]{}\mbox{and}\makebox[2em]{} 
\frac{\phi^{\prime}_{-}(x)}{\phi_{-}(x)}
=-\frac{\kappa}{2}\left\{X-\frac{\epsilon}{4}(p_{1}+p_{2})\right\} 
\end{equation}
the Hamiltonian $H$ is 
\begin{eqnarray}
H&=&-\frac{1}{\kappa}\int dx\triangle\Psi
\nonumber \\
&=&-\frac{1}{\kappa}\left[\Psi^{\prime}\right]^{\infty}_{-\infty}
=\frac{4}{\kappa}\lim_{L\rightarrow \infty}
\left\{\frac{\phi^{\prime}_{+}(L)}{\phi_{+}(L)}
-\frac{\phi^{\prime}_{-}(-L)}{\phi_{-}(-L)}\right\}
\nonumber \\
&=&4X \;\;
\end{eqnarray}
and so (\ref{eqX2}) becomes
\begin{eqnarray}\label{eqH1}
&&(\sqrt{p^{2}_{1}+m^{2}_{1}}-\epsilon\tilde{p}_{2}-H)
(\sqrt{p^{2}_{2}+m^{2}_{2}}+\epsilon\tilde{p}_{1}-H)
\nonumber \\
&& \makebox[5em]{}=(\sqrt{p^{2}_{1}+m^{2}_{1}}-\epsilon\tilde{p}_{1})
(\sqrt{p^{2}_{2}+m^{2}_{2}}+\epsilon\tilde{p}_{2})
e^{\frac{\kappa}{4}\left\{H-\epsilon(\tilde{p}_{1}-\tilde{p}_{2})
\right\}\;|r|} \;\; .
\end{eqnarray}
Eq. (\ref{eqH1}) is the determining equation for the Hamiltonian, whose
solution yields $H$ as a function of $(p_1,p_2,r)$. 

Expanding (\ref{eqH1}) in powers of $\kappa$ yields the perturbative 
solution
\begin{eqnarray}\label{k-exp}
H&=&\sqrt{p^{2}_{1}+m^{2}_{1}}+\sqrt{p^{2}_{2}+m^{2}_{2}}
+\frac{\kappa}{4}(\sqrt{p^{2}_{1}+m^{2}_{1}}\sqrt{p^{2}_{2}+m^{2}_{2}}
-p_{1}p_{2})\;|r|
\nonumber \\
&& +\frac{\kappa\epsilon}{4}(\sqrt{p^{2}_{1}+m^{2}_{1}}\;p_{2}
-p_{1}\sqrt{p^{2}_{2}+m^{2}_{2}})(z_{1}-z_{2})
\nonumber \\
&&+\frac{\kappa^{2}}{2\cdot 4^{2}}\left\{
(\sqrt{p^{2}_{1}+m^{2}_{1}}\sqrt{p^{2}_{2}+m^{2}_{2}}-2p_{1}p_{2})
(\sqrt{p^{2}_{1}+m^{2}_{1}}+\sqrt{p^{2}_{2}+m^{2}_{2}}) \right.
\nonumber \\
&& \makebox[5em]{}\left. +\sqrt{p^{2}_{1}+m^{2}_{1}}\;p^{2}_{2}
+p^{2}_{1}\sqrt{p^{2}_{2}+m^{2}_{2}}\right\}\;r^{2}
\nonumber \\
&& +\frac{\kappa^{2}\epsilon}{2\cdot 4^{2}}\left\{
-2(\sqrt{p^{2}_{1}+m^{2}_{1}}\sqrt{p^{2}_{2}+m^{2}_{2}}-p_{1}p_{2})
(p_{1}-p_{2})+m^{2}_{1}p_{2}-p_{1}m^{2}_{2}\right\}\;|r|(z_{1}-z_{2})
\;\;\cdot\cdot\cdot .
\nonumber \\
\end{eqnarray}
up to ${\cal{O}}(\kappa^2)$.
This is identical with the Hamiltonian derived in the iterative 
method in \cite{OR}.

For the case of $z_{2}<z_{1}$ Eq.(\ref{eqH1}) is
\begin{equation}\label{eqH2}
L_{1}L_{2}=M_{1}M_{2}e^{\frac{\kappa}{4}L_{0}(z_{1}-z_{2})}\;\;.
\end{equation}
Differentiating Eq.(\ref{eqH2}) with respect to $z_{1}$ leads to
\[
\frac{\partial H}{\partial z_{1}}(L_{1}+L_{2})
=\frac{\kappa}{4}L_{1}L_{2}\left(r\frac{\partial H}{\partial z_{1}}
+L_{0}\right)\;\;.
\]
Then we have the canonical equation
\begin{eqnarray}\label{canon1}
\dot{p}_{1}&=&-\frac{\partial H}{\partial z_{1}}
\nonumber \\
&=&\frac{-\frac{\kappa}{4}L_{0}L_{1}L_{2}}
{L_{1}+L_{2}-\frac{\kappa r}{4}L_{1}L_{2}}
\end{eqnarray}
and similarly
\begin{equation}\label{canon2}
\dot{p}_{2}=\frac{\frac{\kappa}{4}L_{0}L_{1}L_{2}}
{L_{1}+L_{2}-\frac{\kappa r}{4}L_{1}L_{2}}\;\;.
\end{equation}
Differentiating Eq.(\ref{eqH2}) with respect to $p_{1}$ leads to
\[
\left(\frac{\partial H}{\partial p_{1}}
-\frac{p_{1}}{\sqrt{p^{2}_{1}+m^{2}_{1}}}\right)L_{1}
+\left(\frac{\partial H}{\partial p_{1}}-\epsilon \right)L_{2}
=L_{1}L_{2}\left\{-\frac{\epsilon}{\sqrt{p^{2}_{1}+m^{2}_{1}}}
+\frac{\kappa r}{4}\left(\frac{\partial H}{\partial p_{1}}-\epsilon
\right)\right\}\;\;.
\]
We have also the canonical equation
\begin{eqnarray}\label{canon3}
\dot{z_{1}}&=&\frac{\partial H}{\partial p_{1}}
\nonumber \\
&=&\epsilon-\frac{\epsilon L_{0}L_{1}}{L_{1}+L_{2}
-\frac{\kappa r}{4}L_{1}L_{2}}
\cdot \frac{1}{\sqrt{p^{2}_{1}+m^{2}_{1}}}
\end{eqnarray}
and similarly
\begin{equation}\label{canon4}
\dot{z}_{2}=-\epsilon+\frac{\epsilon L_{0}L_{2}}
{L_{1}+L_{2}-\frac{\kappa r}{4}L_{1}L_{2}}
\cdot \frac{1}{\sqrt{p^{2}_{2}+m^{2}_{2}}}\;\;.
\end{equation}
It is evident that the Hamiltonian and the total momentum 
$P=p_{1}+p_{2}$ are constants of motion:
\begin{equation}
\dot{H}=0 \qquad \dot{P}=\dot{p}_{1}+\dot{p}_{2}=0\;\;.
\end{equation}

\section{Hamiltonian of two identical particles}

In this section we shall try to solve Eq.(\ref{eqH1}) for a
system of two identical particles. We may choose the center of 
inertia frame with $p_{1}=-p_{2}=p$. Then Eq.(\ref{eqH1}) becomes
\begin{equation}\label{hdef1}
(H-\sqrt{p^{2}+m^{2}}-\epsilon\tilde{p})^{2}
=(\sqrt{p^{2}+m^{2}}-\epsilon\tilde{p})^{2}
e^{\frac{\kappa}{4}(H-2\epsilon\tilde{p})|r|} \quad .
\end{equation}
After setting
\begin{equation}
H=\sqrt{p^{2}+m^{2}}+\epsilon\tilde{p}-\frac{8}{\kappa |r|}Y(p,r)
\end{equation}
eq. (\ref{hdef1}) becomes
\begin{equation}\label{lamb0}
Y^2 e^{2Y} = Z^2 e^{2Z}
\end{equation}
where $Z\equiv \frac{\kappa |r|}{8}(\sqrt{p^{2}+m^{2}}-\epsilon\tilde{p})$.

Equation (\ref{lamb0}) has three solutions shown in Fig.1.  
The trivial solution, $Y=Z$, yields $H=2\epsilon\tilde{p}$, which 
is unphysical because it has no interaction term.  
The second solution (curve A-B-O-C-D) is represented by 
\begin{equation}\label{lamb1}
Y= W(-Z e^{Z}) \qquad Z\leq z_{0}=W^{-1}(e^{-1})
\end{equation}
and the third solution (curve E-F-G) is represented by
\begin{equation}\label{lamb2}
Y= W(Z e^{Z}) \qquad Z<0
\end{equation}
where $W(x)$ is the Lambert $W$-function defined via
\begin{equation}\label{lamb}
y\cdot e^{y}=x \qquad \Longrightarrow \qquad y=W(x)\;\;.
\end{equation}

\hspace*{5cm}--------------------------

\vspace{5mm} 

\hspace*{7cm}Fig.1

\vspace{5mm}

\begin{center}{\it 
Solutions to (\ref{lamb0}). The points B and
C represent the extremal $Z$ and $Y$ values of $W^{-1}(1/e)=0.278$
on the principal branch.
}\end{center}

\hspace*{5cm}--------------------------

In general $W$ is complex and multivalued.  When $x$ is
real, the function has two real branches shown in Fig.2. The branch 
satisfying $-1\leq W(x)$ ( a solid line ) is denoted by $W_{0}(x)$ 
and is referred to as the principal branch.
The branch satisfying $W(x)\leq -1$ ( a broken line ) is denoted by 
$W_{-1}(x)$, and is real-valued only for $-1/e\leq x <0$.
Then for $-1/e\leq x <0$ the function is
double valued. The principal branch is analytic at $x=0$ and has a
derivative singularity at $x=-1/e$ beyond which $W(x)$ becomes complex.
The series expansion of the principal branch is given by
\begin{equation}
W_{0}(x)=\sum_{n=1}^{\infty}\frac{(-n)^{n-1}}{n!}x^{n}\;\;.
\end{equation}

\hspace*{5cm}--------------------------

\vspace{5mm} 

\hspace*{7cm}Fig.2

\vspace{5mm}

\begin{center}{\it 
The Lambert $W$ function
}\end{center}

\hspace*{5cm}--------------------------

The correspondence between the curves in Fig.1 and those in Fig.2 is
\begin{eqnarray}
\mbox{curve A-B-O-C-D} \qquad & \Longleftrightarrow & 
\qquad \mbox{curve P-Q-O-R-O}
\nonumber \\
\mbox{curve E-F-G} \qquad & \Longleftrightarrow & \qquad \mbox{curve O-Q-P}
\nonumber
\end{eqnarray}
Since the physical domain of $Z$ is $Z\geq 0$, the only physical solution
is (\ref{lamb1}), which yields the Hamiltonian 
%
%
\begin{equation}\label{eqmass}
H = \sqrt{p^{2}+ m^2}+ \epsilon p\, \mbox{sgn}(r)
- 8\,\frac{W\left(-\frac{\kappa}{8}
 \left( \! |r|\,\sqrt{p^{2}+ m^2} - {\epsilon{p} r}\, \!  \right) 
\exp\left[{\frac {\kappa}{8}}
 \left( \! |r|\,\sqrt{p^{2}+ m^2} - 
\epsilon{p}\, r \right)\right]\right)}{\kappa|r|} 
\end{equation}
This Hamiltonian 
is exact to infinite order in the gravitational coupling constant. 
We can thus view the whole structure of the theory from the weak 
field to the strong field limits.

The weak field expansion has already been given in the general case 
in (\ref{k-exp}). The small $p$ expansion is
\begin{eqnarray}\label{small-p}
H&=&\frac{m\kappa |r|-8w}{\kappa |r|}+\frac{m\kappa |r|+2m\kappa |r|w
+8w}{m\kappa |r|(1+w)}\;\epsilon\tilde{p}
\nonumber \\
&&-\frac{1}{16}\cdot \frac{-8m\kappa |r|-8m\kappa |r|w^{2}+64w
+m^{2}\kappa^{2} |r|^{2}w}{m^{2}\kappa |r|(1+w)^{3}}\;p^{2}+\cdot\cdot\cdot
\end{eqnarray}
where
\[
w \equiv W\left(-\frac{m\kappa |r|}{8}e^{\frac{m\kappa |r|}{8}}\right)\;\;.
\]
The leading term is simply the mass plus a static gravitational
correction, which is $m^{2}\kappa |r|$ to lowest order in
$\kappa$.
The term linear in $p$ is due to purely to gravity,
since it vanishes in the limit $\kappa \rightarrow 0$. 
The argument of the function $w$ must be larger than $-1/e$. This 
translates into the limit $m\kappa r<W(1/e)$ which means that if 
$m\kappa r$ is sufficiently large, there is no small $p$ expansion -
{\it i.e.} there is some minimum value of $p$ below which the Hamiltonian
is no longer real. This situation is shown in Fig.3. 

\hspace*{5cm}--------------------------

\vspace{5mm} 

\hspace*{7cm}Fig.3 

\vspace{5mm}

\begin{center}{\it 
Hamiltonian as a function of momentum $p>0$ and $\kappa r$ in
units of $m$.
}\end{center}

\hspace*{5cm}--------------------------

The small $m$ expansion (which is the same 
as the large $p$ expansion) is also easily obtained.
For example in the region 
$\;\;p>0, r>0\;\;$ with $\epsilon=1$ we find
\begin{equation}
H=2p+\frac{1}{p}m^{2}+\frac{1}{16}\cdot\frac{-4+\kappa rp}{p^{3}}m^{4}
+\frac{1}{128}\cdot\frac{16-4\kappa rp+\kappa^{2}r^{2}p^{2}}{p^{5}}m^{6}
+\cdot\cdot\cdot \;\;.
\end{equation}
The $\kappa$-independent terms are equivalent to those obtained for
two free relativistic particles of equal mass in the small mass
limit. We see that the effects of gravity modify the Hamiltonian
to include interaction terms whose strength grows with increasing
separation, as one might expect from the basic structure of 
two-dimensional gravity.

The Hamiltonian (\ref{eqmass}) describes the surface in $(r, p, H)$ 
space of all allowed phase-space trajectories. 
Since $H$ is a constant of the motion, a trajectory in the $(r, p)$ plane
is uniquely determined by setting $H=H_{0}$ in (\ref{eqmass}).
However there are two distinct sets of trajectories which
correspond to the two real branches of $W$-function which join 
smoothly onto each other. 

For $H_{0}$ sufficiently small, the Hamiltonian is
given by the principal branch $W_{0}$ and reduces to the Newtonian
limit for small $\kappa$. Fig.4 shows the Hamiltonian for 
small $H_{0}$. The darker surface denotes $H(r, p)=H_{0}$ and
the lighter surface is the Newtonian Hamiltonian.

\hspace*{5cm}--------------------------

\vspace{5mm} 

\hspace*{7cm}Fig.4 

\vspace{5mm}

\begin{center}{\it 
$H$ in the non-relativistic limit compared with the Newtonian
Hamiltonian
}\end{center}

\hspace*{5cm}--------------------------

Once $H_{0}$ becomes sufficiently large, there appears a qualitatively 
new set of trajectories which are not connected with the Newtonian
Hamiltonian in small $\kappa$. Fig.5 demonstrates two branches of
the Hamiltonian in the region $-2< r <-1,\; -2< p <-1$. The whole 
surface of $H$ continues smoothly from one branch to the other. 
This structure is seen in Fig.6 where the slice of $H$
at constant $p \;\;(p=0 )$ is drawn.

\hspace*{5cm}--------------------------

\vspace{5mm} 

\hspace*{7cm}Fig.5 

\vspace{5mm}

\begin{center}{\it 
Both branches of the Hamiltonian
}\end{center}

\hspace*{5cm}--------------------------

\hspace*{5cm}--------------------------

\vspace{5mm} 

\hspace*{7cm}Fig.6 

\vspace{5mm}

\begin{center}{\it 
A slice of $H$ at constant $p$
}\end{center}

\hspace*{5cm}--------------------------

For $H_{0}$ in the intermediate range the constant energy surface 
intersects both branches shown in Fig.7. The trajectory in $(r, p)$
plane moves over both branches. It analytically continues 
from one branch to the other. Fig.8 shows also this structure from
another point of view.

\hspace*{5cm}--------------------------

\vspace{5mm} 

\hspace*{7cm}Fig.7 

\vspace{5mm}

\begin{center}{\it 
A comparison of the Hamiltonian with a surface of constant energy.
The flat black surface corresponds to a value of $H_0=5$. Note that
it intersects both branches of the Hamiltonian.
}\end{center}

\hspace*{5cm}--------------------------

\hspace*{5cm}--------------------------

\vspace{5mm} 

\hspace*{7cm}Fig.8 

\vspace{5mm}

\begin{center}{\it 
A slice of both branches of $H$ at $r=1$. The horizontal line 
corresponds to $H_0=15$ 
}\end{center}

\hspace*{5cm}--------------------------

For a given initial condition the energy $H_{0}$ of the system is 
fixed and the trajectory in $(r, p)$ plane is given as the slice
of $H=H_{0}$ through the 2-dim surface $H(r, p)$ in $(r, p, H)$
phase space. Two characteristic plots are shown in Figs. 9-10
where the corresponding trajectories in the Newtonian theory are
included for comparison. Under time-reversal, the trajectory for 
a given value of $H_{0}$ is obtained by reflection in the $p=0$
axis.

\hspace*{5cm}--------------------------

\vspace{5mm} 

\hspace*{7cm}Fig.9 

\vspace{5mm}

\begin{center}{\it 
Non-relativistic (Newtonian) and relativistic trajectories
for $H_0=2.2$. The undistorted oval is the non-relativistic trajectory.
}\end{center}

\hspace*{5cm}--------------------------

\hspace*{5cm}--------------------------

\vspace{5mm} 

\hspace*{7cm}Fig.10 

\vspace{5mm}

\begin{center}{\it 
Non-relativistic (Newtonian) and relativistic trajectories
for $H_0=3$. The undistorted oval is the non-relativistic trajectory.
}\end{center}

\hspace*{5cm}--------------------------

One of the characteristics of the trajectories is that as 
$H_{0}$ increases the trajectory becomes more $`S'$-shaped. 
Suppose the particles start out at the same place $(r=0)$ with
positive $p$. $r$ will increase and $p$ will slowly decrease.
This continues until maximum separation with some positive
value of $p$, where the velocity $\dot{r}=0$. After that $r$ 
undergoes a rapid decrease, while $p$ is still positive. 
At some value of $r$, $p$ becomes zero and then it goes negative.
The particles continue to be pulled together and $r$ reaches $0$,
where $p$ has its maximum negative value. The particles then 
overshoot the mark and start the reverse motion with interchanged
positions. 

The main reason why the trajectories are $'S'$ shaped is the 
appearance of the $p$-linear term in the Hamiltonian. The canonical
equations (\ref{canon3}) and (\ref{canon4}) (or directly 
the Hamiltonian (\ref{small-p})) leads to
\begin{equation}\label{dot-r}
\dot{r}=\epsilon\;\frac{m\kappa |r|+2m\kappa |r|w+8w}{m\kappa |r|(1+w)}
+(p-\mbox{terms}) \;\;.
\end{equation}
The first term on RHS comes from the $p$-linear term in $H$ and 
$\dot{r}=0$ does not correspond to $p=0$.  
This relation between $\dot{r}$ and $p$ resembles the relation in 
the theory with charged particles
\[
{\bf p}=\frac{\dot{\bf r}}{\sqrt{1-\dot{\bf r}^{2}}}+e{\bf A}\;\;.
\]
In this sense the first term on RHS of (\ref{dot-r}) can be said 
to be purely gravitational.

\section{The unequal mass Hamiltonian}

For unequal masses  we set
\begin{equation}
H=\frac{\sqrt{p^{2}_{1}+m^{2}_{1}}+\sqrt{p^{2}_{2}+m^{2}_{2}}}{2}
+\frac{\epsilon}{2}(\tilde{p}_{1}-\tilde{p}_{2}) - \frac{8}{\kappa|r|}Y
\end{equation}
so that (\ref{eqH1}) becomes
\begin{equation}\label{glamb0}
(Y^2-D^2) e^{2Y} = (S^2-D^2) e^{2S}
\end{equation}
where $S=\frac{\kappa |r|}{16}(\tilde{M}_1+\tilde{M}_2)$ and
$D=\frac{\kappa |r|}{16}(\tilde{M}_1-\tilde{M}_2)$ with $\tilde{M}_1
\equiv \sqrt{p^{2}_{1}+m^{2}_{1}}-\epsilon\tilde{p}_1$ and $\tilde{M}_2
\equiv \sqrt{p^{2}_{2}+m^{2}_{2}}+\epsilon\tilde{p}_2$. 
For equal masses, $S=Z$, $D=0$ and (\ref{glamb0}) reduces to (\ref{lamb0}).
Solving (\ref{glamb0}) for $Y$ in terms of $S$ and $D$ yields the
Hamiltonian in the unequal mass case.

To obtain the solution, consider the equation
\begin{equation}\label{glamb}
(y^2-a^2) e^{2y} = (x^2-a^2) e^{2x} \qquad a>0\;\;.
\end{equation}
This equation also has three solutions shown in Fig.11 : 
the trivial solution $y=x$, the curve H-I-J-K-L denoted by 
${\cal W}(x;a)$ and the curve S-T-U denoted by $\bar{\cal W}(x;a)$.
To our knowledge, discussion of the functions ${\cal W}(x;a)$ and
$\bar{\cal W}(x;a)$ have never appeared in the literature.  
We shall refer to ${\cal W}$ as the generalized
Lambert function since $\lim_{a\to 0}{\cal W}(x;a) = W(-x e^x)$.
In general ${\cal W}$ is also complex and multivalued, and when
$x$ is real, the function has two real branches shown in Fig.11. 
The principal branch is analytic at $x=0$ and has a
derivative singularity at $x={\cal W}^{-1}\left(-\frac{1}{2}(1+\sqrt{1+4a^2})
\right)$ beyond which it becomes complex. The other branch 
satisfies ${\cal W} < -\frac{1}{2}(1+\sqrt{1+4a^2})$ and joins
smoothly onto the first branch. The full function is double valued
for $ a < x < {\cal W}^{-1}(-\frac{1}{2}(1+\sqrt{1+4a^2}))$. 
The third solution $\bar{\cal W}(x;a)$ is a generalization of 
$W(xe^{x})$ in the region $x<0$.

\hspace*{5cm}--------------------------

\vspace{5mm} 

\hspace*{7cm}Fig.11

\vspace{5mm}

\begin{center}{\it 
A plot of solutions to eq. (82).  The curve HIJKL is 
the Generalized Lambert function.
}\end{center}

\hspace*{5cm}--------------------------

As in the equal mass case, the trivial solution  $Y=S$ 
again yields the unphysical Hamiltonian 
$H=\epsilon(\tilde{p}_1-\tilde{p}_2)$. Since the physical domain of $S$
is $S\geq 0$, the only physical solution is
\begin{equation}\label{glamb1}
Y= {\cal W}(S;D) 
\end{equation}
which leads to the Hamiltonian
\begin{equation}
H=\frac{\sqrt{p^{2}_{1}+m^{2}_{1}}+\sqrt{p^{2}_{2}+m^{2}_{2}}}{2}
+\frac{\epsilon}{2}(\tilde{p}_{1}-\tilde{p}_{2}) 
- 8\frac{\kappa}{|r|}{\cal W}[\frac{\kappa |r|}{16}(\tilde{M}_1+\tilde{M}_2);
\frac{\kappa |r|}{16}(\tilde{M}_1-\tilde{M}_2)] \quad .
\end{equation}
The expansion in $\kappa$ for this Hamiltonian is given by 
(\ref{k-exp}).

Choosing also the center of inertia frame with $p_{1}=-p_{2}=p$
and setting $m=m_{2}/m_{1}$, we shall look at the trajectories.
First, take a value for $H_{0}$ just above the minimal (rest-mass)
value of $1+m$ and compare this to Newtonian theory in Fig. 12. 

\hspace*{5cm}--------------------------

\vspace{5mm} 

\hspace*{4cm}Fig. 12 

\vspace{5mm}

\begin{center}{\it 
Non-relativistic (Newtonian) and relativistic trajectories
for $H_0=2.01$ in the unequal mass case with $m=1$.
The undistorted oval is the non-relativistic trajectory.
}\end{center}

\hspace*{5cm}--------------------------

\noindent
The trajectory is almost exactly the same as Newtonian theory, 
since it is the equal mass case. For larger $m$ the separation
between particles cannot get to be very large and the trajectory
becomes more compact as shown in Fig. 13.

\hspace*{5cm}--------------------------

\vspace{5mm} 

\hspace*{4cm}Fig.13 

\vspace{5mm}

\begin{center}{\it 
Non-relativistic (Newtonian) and relativistic trajectories
for $H_0=6.01$ in the unequal mass case with $m=5$.
The narrow oval in the middle is the relativistic trajectory.
}\end{center}

\hspace*{5cm}--------------------------

\noindent
The trajectories for smaller values of $m$ are shown in Fig. 14, 
where the innermost line is the $m=0.9$ case and the outermost
is the $m=0.1$ case.

\hspace*{5cm}--------------------------

\vspace{5mm} 

\hspace*{5cm}Fig.14 

\vspace{5mm} 

\begin{center}{\it 
Relativistic trajectories for several values of $m$,
where  $H_0=2$. 
}\end{center}

\hspace*{5cm}--------------------------

\noindent
Finally in Fig.15 the trajectories of different values of $m$
both large and small are compared.

\hspace*{5cm}--------------------------

\vspace{5mm} 

\hspace*{4cm}Fig.15 

\vspace{5mm}

\begin{center}{\it 
Relativistic trajectories for several values of $m$,
where  $H_0=4$. 
}\end{center}

\hspace*{5cm}--------------------------

\section{Solution of the metric tensor}

To determine the Hamiltonian and derive the canonical equations of
motion, we had only to solve the constraints (\ref{feq3}) and 
(\ref{feq4}) of the system of the field equations (\ref{feq1})-
(\ref{feq8}). In this section we shall solve the remaining 
equations to determine the metric and to confirm directly the 
consistency of Euler-Lagrange equations (\ref{feq7}) and (\ref{feq8})
with the canonical equations derived from the Hamiltonian, though 
formal proof of the consistency was already given in \cite{OR}.

Under the coordinate conditions (\ref{cc2}) the field equations 
(\ref{feq1}), (\ref{feq2}), (\ref{feq5}) and (\ref{feq6}) become
\begin{eqnarray}\label{feq10}
\dot{\pi}&+&N_{0}\left\{\frac{3\kappa}{2}\pi^{2}
+\frac{1}{8\kappa}(\Psi^{\prime})^{2}
-\sum_{a}\frac{p^{2}_{a}}{2\sqrt{p^{2}_{a}+m^{2}_{a}}}
\;\delta(x-z_{a}(x^{0}))\right\}
\nonumber \\
&+&N_{1}\left\{\pi^{\prime}
+\sum_{a}p_{a}\;\delta(x-z_{a}(x^{0}))\right\}
+\frac{1}{2\kappa}N^{\prime}_{0}\Psi^{\prime}
+N^{\prime}_{1}\pi=0
\end{eqnarray}
\begin{equation}\label{feq20}
\kappa\pi N_{0}+N^{\prime}_{1}=0
\end{equation}
\begin{equation}\label{feq50}
\partial_{1}(\frac{1}{2}N_{0}\Psi^{\prime}+N^{\prime}_{0})=0 
\end{equation}
\begin{equation}\label{feq60}
\dot{\Psi}=-2\kappa\pi N_{0}+N_{1}\Psi^{\prime}\;\;.
\end{equation}

In the following we shall carry out our calculations assuming 
$z_{2}<z_{1}$ -- the case $z_{1}<z_{2}$ is completely
analogous and will not be presented.
The solution to (\ref{feq50}) is 
\begin{equation}
N_{0}=Ae^{-\frac{1}{2}\Psi}=A\phi^{2}=\left\{
\begin{array}{ll}
A\phi^{2}_{+} & \makebox[3em]{}(+)\mbox{region} \\
A\phi^{2}_{0} & \makebox[3em]{}(0)\mbox{region} \\
A\phi^{2}_{-} & \makebox[3em]{}(-)\mbox{region}
\end{array}
\right.
\end{equation}
where $A$ is an integration constant and $\phi_{\pm}$ and $\phi_{0}$
are given in (\ref{phi+}), (\ref{phi-}) and (\ref{phi0}).
Eq.(\ref{feq20}) becomes
\begin{equation}
N^{\prime}_{1}=-\kappa A\chi^{\prime}\phi^{2}\;\;.
\end{equation}
The solution in each region is

\vspace{5mm}
\noindent
(+) region :
\begin{equation}\label{N1plus}
N_{1(+)}=\epsilon\left\{A\frac{L_{1}}{M_{1}}
e^{\frac{\kappa}{4}L_{+}(x-z_{1})}-1\right\}
=\epsilon\left(A\phi^{2}_{+}-1\right)
\end{equation}

\vspace{5mm}
\noindent
(-) region :
\begin{equation}\label{N1minus}
N_{1(-)}=-\epsilon\left\{A\frac{L_{2}}{M_{2}}
e^{-\frac{\kappa}{4}L_{-}(x-z_{2})}-1\right\}
=-\epsilon\left(A\phi^{2}_{-}-1\right)
\end{equation}

\vspace{5mm}
\noindent
(0) region :
\begin{equation}
N_{1(0)}=\epsilon A\frac{L_{1}L_{2}}{L^{2}_{0}}\left\{
\frac{L_{2}}{M_{1}}e^{\frac{\kappa}{4}L_{0}(x-z_{1})}
-\frac{L_{1}}{M_{2}}e^{-\frac{\kappa}{4}L_{0}(x-z_{2})}\right\}
+\frac{\kappa\epsilon}{2}A\frac{L_{1}L_{2}}{L_{0}}x+\epsilon C_{0}
\end{equation}
where we chose the integration constants in $N_{1(+)}$ and $N_{1(-)}$
to be $(-\epsilon)$ and $\epsilon$, respectively. 
In general we should respectively
replace these with arbitrary constants $C_{+}$ and 
$C_{-}$ appear in (\ref{N1plus}) and 
(\ref{N1minus}). However
a lengthy calculation reveals that they are equal to $(-\epsilon)$
and $\epsilon$, respectively. For simplicity we shall
set $C_{+}=-\epsilon$ and $C_{-}=\epsilon$
from the outset. In deriving $N_{1(0)}$ we used (\ref{eqH2}).

The continuity condition at $x=z_{1}$
\begin{equation}
N_{1(+)}(z_{1})=N_{1(0)}(z_{1})
\end{equation}
leads to
\begin{equation}\label{C01}
C_{0}=-1+A\left\{\frac{L_{1}}{M_{1}}-\frac{L_{1}(L_{2}-M_{1})}{L_{0}M_{1}}
-\frac{\kappa}{2}\frac{L_{1}L_{2}}{L_{0}}z_{1}\right\}
\end{equation}
where (\ref{eqH2}) and the relation $L_{2}+M_{1}=L_{0}$ are used.
The continuity condition at $x=z_{2}$
\begin{equation} 
N_{1(-)}(z_{2})=N_{1(0)}(z_{2})
\end{equation}
similarly leads to
\begin{equation}\label{C02}
C_{0}=1-A\left\{\frac{L_{2}}{M_{2}}-\frac{L_{2}(L_{1}-M_{2})}{L_{0}M_{2}}
+\frac{\kappa}{2}\frac{L_{1}L_{2}}{L_{0}}z_{2}\right\}
\end{equation}
{}From the consistency of (\ref{C01}) and (\ref{C02}) 
the constant $A$ is determined as
\begin{equation}
A=\frac{L_{0}}{L_{1}+L_{2}-\frac{\kappa r}{4}L_{1}L_{2}}
\end{equation}
and
\begin{equation}
C_{0}=\frac{M_{1}-M_{2}-\frac{\kappa}{4}(z_{1}+z_{2})L_{1}L_{2}}
{L_{1}+L_{2}-\frac{\kappa r}{4}L_{1}L_{2}} \;\;.
\end{equation}

\vspace{5mm}
Now we are ready to check (\ref{feq10}).
First we treat three regions (+), (0) and (-) separately, and next
consider the matching conditions at $x=z_{1}, z_{2}$.

For (+) and (-) regions it is straightforward to show that the LHS of 
(\ref{feq10}) vanishes, by substituting the explicit solutions of
$\pi=\chi^{\prime}, \Psi^{\prime}=-4\phi^{\prime}/\phi, N_{0}$
and $N_{1}$. 
For the (0) region the solutions of the metric and the dilaton field
are 
\begin{eqnarray}
\pi &=&-\frac{\epsilon}{4}L_{0}, \makebox[2em]{} \pi^{\prime}=0
\nonumber \\
\dot{\pi}&=&-\frac{\epsilon}{4}\dot{L}_{0}
=-\frac{\epsilon}{4}(4\dot{X}-\epsilon\dot{p}_{1}
+\epsilon\dot{p}_{2})=\frac{1}{2}\dot{p}_{1}
\nonumber \\
\Psi^{\prime}&=&-4\frac{\phi^{\prime}_{0}}{\phi_{0}}
\nonumber \\
N_{0}&=&A\phi^{2}_{0}, \makebox[2em]{} N^{\prime}_{0}=
2A\phi_{0}\phi^{\prime}_{0}
\nonumber \\
N^{\prime}_{1}&=&\frac{\kappa\epsilon}{4}L_{0}A\phi^{2}_{0}\;\;.
\nonumber 
\end{eqnarray}
The LHS of (\ref{feq10}) becomes
\begin{eqnarray}
\mbox{LHS of (\ref{feq10})} 
&=&\frac{1}{2}\dot{p}_{1}+\frac{\kappa A}{32}(L_{0}\phi_{0})^{2}
-\frac{2A}{\kappa}(\phi^{\prime}_{0})^{2}
\nonumber \\ 
&=&\frac{1}{2}\dot{p}_{1}+\frac{\kappa}{8}\frac{L_{0}L_{1}L_{2}}
{L_{1}+L_{2}-\frac{\kappa r}{8}L_{1}L_{2}}
\nonumber
\end{eqnarray}
which vanishes due to (\ref{canon1}).

Consider next the $\delta$-function part at $x=z_{1}$.
Since
\begin{eqnarray}
\pi&=&\chi^{\prime}=-\frac{1}{4}\left\{p_{1}\mid x-z_{1}\mid^{\prime}
+p_{2}\mid x-z_{2}\mid^{\prime}\right\}-\epsilon X
\nonumber \\
\pi^{\prime}&=&-\frac{1}{2}\left\{p_{1}\delta(x-z_{1})
+p_{2}\delta(x-z_{2})\right\}
\nonumber \\
\dot{\pi}&=&\frac{1}{2}\left\{p_{1}\dot{z}_{1}\delta(x-z_{1})
+p_{2}\dot{z}_{2}\delta(x-z_{2})\right\}
-\frac{1}{4}\left\{\dot{p}_{1}\mid x-z_{1}\mid^{\prime}
+\dot{p}_{2}\mid x-z_{2}\mid^{\prime}\right\}
\nonumber
\end{eqnarray}
we have
\begin{eqnarray}
\lefteqn{[\delta-\mbox{function part of (11) at}\makebox[0.5em]{} x=z_{1}]}
\nonumber \\
&=&\frac{1}{2}p_{1}\dot{z}_{1}\delta(x-z_{1})
-\frac{1}{2}N_{0}(z_{1})\frac{p^{2}_{1}}{\sqrt{p^{2}_{1}+m^{2}_{1}}}
\delta(x-z_{1})+\frac{1}{2}N_{1}(z_{1})p_{1}\delta(x-z_{1})
\nonumber \\
&=&\frac{1}{2}p_{1}\delta(x-z_{1})\left\{\dot{z}_{1}
+\frac{\epsilon L_{0}L_{1}}{L_{1}+L_{2}-\frac{\kappa r}{4}L_{1}L_{2}}\cdot
\frac{1}{\sqrt{p^{2}_{1}+m^{2}_{1}}}-\epsilon\right\}\;\;.
\nonumber 
\end{eqnarray}
which also vanishes due to
the canonical equation (\ref{canon3}).
Similarly, the $\delta$-function part of (11) at $x=z_{2}$ is zero.
We thus conclude that (\ref{feq10}) is satisfied exactly.

As we investigated in the iterative method \cite{OR}, for the consistency
of (\ref{feq60}) we need to introduce a $x$-independent function $f(t)$ 
into $\Psi$
\begin{equation}
\Psi=-4\mbox{log}\mid \phi\mid+f(t) \;\;.
\end{equation}
Since in the system of the original equations (\ref{feq1})-(\ref{feq8})
all other equations except (\ref{feq5}) contain only spatial 
derivatives of $\Psi$, $f(t)$ does not contribute to either the 
Hamiltonian or to the equations of motion. Eq. (\ref{feq60}) becomes
\begin{equation}\label{eq-f}
-4\frac{\dot{\phi}}{\phi}+\dot{f}(t)+2\kappa\pi N_{0}
+4\frac{\phi^{\prime}}{\phi}N_{1}=0
\end{equation}
We must check this equation in the three regions separately, with
$f(t)$ common to all regions.  

For the (+) region, after substituting the solutions of $\phi_{+},
\pi, N_{0(+)}$ and $N_{1(+)}$, (\ref{eq-f}) yields
\begin{equation}
\dot{f}_{+}(t)=2\frac{\dot{L}_{1}M_{1}-L_{1}\dot{M}_{1}}{L_{1}M_{1}}
-\frac{\kappa}{2}L_{+}(\dot{z}_{1}-\epsilon)
\end{equation}
This ensures that $f_{+}$ is $x$-independent.
Using the canonical equations (\ref{canon1}), (\ref{canon2}), 
(\ref{canon3}) and (\ref{canon4}) we get
\begin{equation}\label{fplusd}
\dot{f}_{+}(t)=\frac{\frac{\kappa}{2}L_{0}}
{L_{1}+L_{2}-\frac{\kappa r}{4}L_{1}L_{2}}\left\{
L_{1}\left(\epsilon+\frac{p_{1}}{\sqrt{p^{2}_{1}+m^{2}_{1}}}\right)
+L_{2}\left(\epsilon-\frac{p_{2}}{\sqrt{p^{2}_{2}+m^{2}_{2}}}\right)\right\}
\end{equation}
For the (-) region the calculation is quite analogous to the above and 
leads to $ \dot{f}_{-}(t)=\dot{f}_{+}(t)$.
For the (0) region the calculation is rather lengthy and complicated, 
especially for $\dot{\phi}_{0}$, which is expressed as
\begin{eqnarray}
\dot{\phi}_{0}
&=&\left(-\frac{\dot{L}_{0}}{L_{0}}+\frac{\dot{L}_{1}}{2L_{1}}
+\frac{\dot{L}_{2}}{2L_{2}}+\frac{\kappa\epsilon}{8}\cdot
\frac{\frac{\kappa r}{4}L_{0}L_{1}L_{2}}{L_{1}+L_{2}-\frac{\kappa r}{4}
L_{1}L_{2}}\right)\phi_{0}
\nonumber \\
&&+\frac{\dot{L}_{0}}{L_{0}}x\phi^{\prime}_{0}
+\frac{\epsilon(M_{1}-M_{2})}{L_{1}+L_{2}-\frac{\kappa r}{4}L_{1}L_{2}}
\phi^{\prime}_{0}
\nonumber \\
&&-\frac{\kappa}{8}\cdot \frac{\dot{L}_{0}}{L_{0}}(L_{1}L_{2})^{1/2}
\left\{\left(\frac{L_{2}}{M_{1}}\right)^{1/2}z_{1}
e^{\frac{\kappa}{8}L_{0}(x-z_{1})}
-\left(\frac{L_{1}}{M_{2}}\right)^{1/2}z_{2}e^{-\frac{\kappa}{8}L_{0}
(x-z_{2})}\right\}
\nonumber
\end{eqnarray}
Substitution of the expressions of $\phi_{0}, \phi^{\prime}_{0},
\dot{\phi}_{0}, \pi, N_{0}$ and $N_{1(0)}$ into (\ref{eq-f}) 
in the (0)-region leads to
\begin{eqnarray}
\dot{f}_{0}(t)&=&4\left(-\frac{\dot{L}_{0}}{L_{0}}
+\frac{\dot{L}_{1}}{2L_{1}}+\frac{\dot{L}_{2}}{2L_{2}}\right)
+2\kappa\epsilon A\frac{L_{1}L_{2}}{L_{0}}
\nonumber \\
&=&\frac{\frac{\kappa}{2}L_{0}}
{L_{1}+L_{2}-\frac{\kappa r}{4}L_{1}L_{2}}
\left\{L_{1}\left(\epsilon+\frac{p_{1}}{\sqrt{p^{2}_{1}+m^{2}_{1}}}\right)
+L_{2}\left(\epsilon-\frac{p_{2}}{\sqrt{p^{2}_{2}+m^{2}_{2}}}\right)\right\}
\end{eqnarray}
which is equivalent to (\ref{fplusd}).
Hence $f(t)$ is common in all regions
\begin{equation}
\dot{f}_{+}(t)=\dot{f}_{-}(t)=\dot{f}_{0}(t) \;\;
\end{equation}
and the solution is self-consistent.

Finally we shall directly check the Euler-Lagrange equations
(\ref{feq7}) and (\ref{feq8}) which  under the coordinate
conditions (\ref{cc2}) become
\begin{equation}\label{EL1}
\dot{z}_{a}-N_{0}(z_{a})\frac{p_{a}}{\sqrt{p^{2}_{a}+m^{2}_{a}}}
+N_{1}(z_{a})=0
\end{equation}
\begin{equation}\label{EL2}
\dot{p}_{a}+\frac{\partial N_{0}}{\partial z_{a}}\sqrt{p^{2}_{a}+m^{2}_{a}}
-\frac{\partial N_{1}}{\partial z_{a}}p_{a}=0 \;\;.
\end{equation}
Since $N_{0}$ and $N_{1}$ are continuous at $x=z_{1}, z_{2}$, we have
\begin{eqnarray}
N_{0}(z_{1})&=&A\phi^{2}_{+}(z_{1})=\frac{L_{0}}
{L_{1}+L_{2}-\frac{\kappa r}{4}L_{1}L_{2}}\cdot \frac{L_{1}}{M_{1}}
\nonumber \\
N_{1}(z_{1})&=&\epsilon\left(\frac{L_{0}}
{L_{1}+L_{2}-\frac{\kappa r}{4}L_{1}L_{2}}\cdot \frac{L_{1}}{M_{1}}-1\right)
\;\;. \nonumber
\end{eqnarray}
Then for particle 1, say, (\ref{EL1}) is
\begin{equation}
\dot{z}_{1}
=\epsilon-\frac{\epsilon L_{0}L_{1}}{L_{1}+L_{2}
-\frac{\kappa r}{4}L_{1}L_{2}}
\cdot \frac{1}{\sqrt{p^{2}_{1}+m^{2}_{1}}}
\end{equation}
This is identical with the canonical equation (\ref{canon3}).

On the other hand, $\partial N_{0}/\partial x$ and $\partial N_{1}
/\partial x$ are discontinuous at $x=z_{1}, z_{2}$. The natural
definition of $\partial N_{0}/\partial z_{1}$ is 
\begin{eqnarray}
\frac{\partial N_{0}}{\partial z_{1}}&\equiv &\frac{1}{2}\left\{
\left.\frac{\partial N_{0}}{\partial x}\right|_{x=z_{1}+0}
+\left.\frac{\partial N_{0}}{\partial x}\right|_{x=z_{1}-0}\right\}
\\
&=&\frac{\kappa}{8}A\frac{L_{1}}{M_{1}}\left\{L_{+}+L_{2}-M_{1}\right\}
\end{eqnarray}
and similarly
\begin{eqnarray}
\frac{\partial N_{1}}{\partial z_{1}}&\equiv &\frac{1}{2}\left\{
\left.\frac{\partial N_{1}}{\partial x}\right|_{x=z_{1}+0}
+\left.\frac{\partial N_{1}}{\partial x}\right|_{x=z_{1}-0}\right\}
\\
&=&\frac{\kappa\epsilon}{8}A\frac{L_{1}}{M_{1}}(L_{+}+L_{0}) \;\;.
\end{eqnarray}
For  particle 1, (\ref{EL2}) is
\begin{equation}
\dot{p}_{1}
=\frac{-\frac{\kappa}{4}L_{0}L_{1}L_{2}}{L_{1}+L_{2}-\frac{\kappa r}{4}
L_{1}L_{2}}
\end{equation}
which is identical to (\ref{canon1}). For  particle 2, 
(\ref{EL1}) and (\ref{EL2}) also reproduce the canonical equations
(\ref{canon2}) and (\ref{canon4}).

Thus the consistency of the solution has been completely proved.

\section{Test particle approximation}

As an interesting limiting case of (\ref{eqH1}) let us try to get the 
Hamiltonian in the test-particle approximation. 

Setting particle 1 to be a test particle $\mu$  and particle 2 to be
a static source $m$ at the origin, namely,
\begin{eqnarray}
&&z_{1}=z, \makebox[4em]{} m_{1}=\mu, \makebox[4em]{} p_{1}=p,
\makebox[4em]{} \tilde{p}_{1}=\tilde{p}=p\frac{z}{\mid z\mid}
\nonumber \\
&&z_{2}=0, \makebox[4em]{} m_{2}=m, \makebox[4em]{} p_{2}=0,
\makebox[4em]{} \tilde{p}_{2}=0
\nonumber 
\end{eqnarray}
the defining equation (\ref{eqH1}) for the Hamiltonian becomes
\begin{equation}
(\sqrt{p^{2}+\mu^{2}}-H)(m+\epsilon\tilde{p}-H)=
(\sqrt{p^{2}+\mu^{2}}-\epsilon\tilde{p})m\;
e^{\frac{\kappa}{4}(H-\epsilon\tilde{p})\mid z\mid} \;\;.
\end{equation}
Expanding $H$ in a power series in $\sqrt{p^{2}+\mu^{2}}$ and 
$\epsilon\tilde{p}$ and taking only the linear terms  we obtain
\begin{equation}
H=m+\sqrt{p^{2}+\mu^{2}}\;e^{\frac{\kappa m}{4}\mid z\mid}
-\epsilon p\;\frac{z}{\mid z\mid}\left(e^{\frac{\kappa m}{4}\mid z\mid}-1
\right) \;\;.
\end{equation}
for the Hamiltonian in the test particle approximation.
This Hamiltonian is expressed in terms of the metric tensor of the static
source as
\begin{equation}\label{testham}
H=m+\sqrt{p^{2}+\mu^{2}}\;N_{0}(z)-pN_{1}(z)
\end{equation}
where
\begin{equation}\label{Ntest}
N_{0}=e^{\frac{\kappa m}{4}\mid z\mid} \qquad
N_{1}=\epsilon\;\frac{z}{\mid z\mid}
\left(e^{\frac{\kappa m}{4}\mid z\mid}-1\right) \;\;.
\end{equation}

The canonical equations are
\begin{equation}\label{test1}
\dot{z}=\frac{p}{\sqrt{p^{2}+\mu^{2}}}N_{0}-N_{1}
\end{equation}
and
\begin{equation}
\dot{p}=-\sqrt{p^{2}+\mu^{2}}\frac{\partial N_{0}}{\partial z}
-p\frac{\partial N_{1}}{\partial z}=0 \;\;.
\end{equation}
Eq.(\ref{test1}) is inversely solved as
\begin{equation}\label{testp1}
p= \frac{\mu(N_{1}+\dot{z})}{\left[N^{2}_{0}-(N_{1}
+\dot{z})^{2}\right]^{\frac{1}{2}}} \;\;.
\end{equation}
We set the initial condition
\begin{equation}
z=0  \qquad \dot{z}=v_{0} \makebox[2em]{}\mbox{at}
\makebox[2em]{}t=0 \;\;.
\end{equation}
Then the initial momentum $p(0)$ and the total energy $H_{0}$ 
are
\begin{equation}\label{init}
p(0)=\frac{\mu v_{0}}{\sqrt{1-v^{2}_{0}}} \qquad 
H_{0}=m+\sqrt{p(0)^{2}+\mu^{2}}
\end{equation}
{}From (\ref{testham}) and (\ref{init}) $p$ is given by
\begin{equation}\label{testp2}
p=\frac{1}{N^{2}_{0}-N^{2}_{1}}\left\{\sqrt{p(0)^{2}+\mu^{2}}N_{1}
\pm N_{0}\sqrt{p(0)^{2}+\mu^{2}-(N^{2}_{0}-N^{2}_{1})\mu^{2}}\right\}
\;\;.
\end{equation}

We can draw a trajectory in phase space, an example of which is shown 
in Fig.16. The trajectory is again $'S'$ shaped due to  
relativistic gravitational effects.

\hspace*{5cm}--------------------------

\vspace{5mm} 

\hspace*{7cm}Fig.16

\vspace{5mm}

\begin{center}{\it 
Relativistic trajectory for a test particle compared to the
Newtonian case.
}\end{center}

\hspace*{5cm}--------------------------

{}From (\ref{testp1}) and (\ref{testp2}) we get
\begin{equation}
\dot{z}=\frac{N_{0}\left[N_{1}+ 
N_{0}\sqrt{1-(N^{2}_{0}-N^{2}_{1})(1-v^{2}_{0})}\;\right]}
{\sqrt{(N^{2}_{0}-N^{2}_{1})^{2}(1-v^{2}_{0})+\left[N_{1}+ 
N_{0}\sqrt{1-(N^{2}_{0}-N^{2}_{1})(1-v^{2}_{0})}\;\right]^{2}}}-N_{1}
\;\;.
\end{equation}
Denoting the RHS as $G(z)$, the solution is given by
\begin{equation}
t=\int ^{z}_{0}\frac{dz}{G(z)} \;\;.
\end{equation}
To lowest order in $\kappa$ we obtain
\begin{equation}
G(z)= \sqrt{v^{2}_{0}-\frac{\kappa m}{2}|z|}
\end{equation}
and
\begin{equation}
z=-\frac{\kappa m}{8}t^{2}+v_{0}t 
\end{equation}
which is the Newtonian motion of a test body in (1+1) dimensions.

We shall add one comment on the form of the line element. In our 
canonical reduction we chose the coordinate conditions (\ref{cc2}),
under which the line element of space-time is
\begin{equation}
ds^{2}=-(N^{2}_{0}-N^{2}_{1})dt^{2}+2N_{1}dtdx +dx^{2}\;\;.
\end{equation}
For the case of a single static source with $N_{0}$ and $N_{1}$
given by (\ref{Ntest}), we find the coordinate transformations
\begin{eqnarray}
\tilde{t} &=& t-\epsilon |x| +\frac{2\epsilon}{\kappa m}\mbox{log}
|2e^{\frac{\kappa m}{4} |x| }-1|
\\
\tilde{x} &=& \frac{4}{\kappa m}\cdot\frac{x}{ |x| }
\left(e^{\frac{\kappa m}{4} |x| }-1\right)
\end{eqnarray}
which leads to the line element \cite{MST}
\begin{equation}
ds^{2}=-\alpha(\tilde{x})d\tilde{t}^{2}
+\frac{1}{\alpha(\tilde{x})}d\tilde{x}^{2}
\end{equation}
with
\begin{equation}
\alpha(\tilde{x})=1+\frac{\kappa m}{2} |\tilde{x}| \;\;.
\end{equation}
In this coordinate frame the Hamiltonian for the test particle
is
\begin{equation}
H(\tilde{z}, \tilde{p})=\sqrt{\alpha(\tilde{z})\mu^{2}
+\alpha(\tilde{z})^{2}p^{2}}+m \;\;.
\end{equation}

\section{Correspondence with Newtonian gravity in $(d+1)$ dimensions} 

In this section we illustrate how a Newtonian limit generically arises in
the $(1+1)$ dimensional theory we consider.  We compare this with
the emergence of a Newtonian limit in $(d+1)$ dimensions.
We shall compute the Newtonian limit(s) by considering the one graviton 
exchange potential (keeping in mind that there are no propagating gravitons
in two spacetime dimensions).

We begin by extending the theory in (\ref{act1}) to $d+1=n$ dimensions and 
coupling $N$ scalar fields, which yields
\begin{equation}\label{Lag1}
L=\frac{2}{\kappa^{2}}\sqrt{-g}\left\{\Psi R
+\frac{1}{2}g^{\mu\nu}\nabla_{\mu}\Psi\nabla_{\nu}\Psi\right\}
-\frac{1}{2}\sum_{a}\sqrt{-g}\left(g^{\mu\nu}\phi_{a,\mu}\phi_{a,\nu}
+m^{2}_{a}\phi^{2}_{a}\right)
\end{equation}
where $\kappa^{2}=32\pi G$. Defining the graviton field $h_{\mu\nu}$ 
and the dilaton field $\psi$ via
\begin{equation}
g_{\mu\nu}=\eta_{\mu\nu}+\kappa h_{\mu\nu} 
\qquad
\Psi=1+\kappa\psi\;\;
\end{equation}
gives
\begin{eqnarray}\label{Lag2}
L_{0}&=&-\frac{1}{2}\left\{\partial_{\lambda}h^{\mu\nu}\partial^{\lambda}
h_{\mu\nu}-\partial^{\lambda}h^{\mu}_{\;\;\mu}
\partial_{\lambda}h^{\nu}_{\;\;\nu}
-2\partial_{\mu}h^{\mu\nu}\partial^{\lambda}h_{\lambda\nu}
+2\partial_{\mu}h^{\mu\nu}\partial_{\nu}h^{\lambda}_{\;\;\lambda}\right\}
\nonumber \\
&&-2\left(\partial^{\nu}h_{\mu\nu}
-\partial_{\mu}h^{\nu}_{\;\nu}\right)\partial^{\mu}\psi 
+\partial^{\mu}\psi\partial_{\mu}\psi
-\frac{1}{2}\sum_{a}\left(\phi^{,\;\mu}_{a}\phi_{a,\;\mu}
+m^{2}_{a}\phi^{2}_{a}\right)\;\;.
\end{eqnarray}
for the free Lagrangian density following from (\ref{Lag1}).
Redefining the dilaton field 
\begin{equation}
\tilde{\psi}\equiv \psi+h^{\mu}_{\;\mu}-\frac{\partial^{\mu}\partial^{\nu}}
{\Box}h_{\mu\nu}
\end{equation}
allows us to express (\ref{Lag2}) as
\begin{eqnarray}\label{Lag3}
L_{0}&=&-\frac{1}{2}\left\{\partial_{\lambda}h^{\mu\nu}\partial^{\lambda}
h_{\mu\nu}-\partial^{\lambda}h^{\mu}_{\;\;\mu}
\partial_{\lambda}h^{\nu}_{\;\;\nu}
-2\partial_{\mu}h^{\mu\nu}\partial^{\lambda}h_{\lambda\nu}
+2\partial_{\mu}h^{\mu\nu}\partial_{\nu}h^{\lambda}_{\;\;\lambda}\right\}
\nonumber \\
&&-\left(\partial^{\nu}h_{\mu\nu}-\partial_{\mu}h^{\nu}_{\;\nu}\right)
\left(\partial^{\lambda}h^{\mu}_{\;\lambda}
-\partial^{\mu}h^{\lambda}_{\;\lambda}\right)
+\partial_{\mu}\tilde{\psi}\partial^{\mu}\tilde{\psi}
-\frac{1}{2}\sum_{a}\left(\phi^{,\;\mu}_{a}\phi_{a,\;\mu}
+m^{2}_{a}\phi^{2}_{a}\right)\;\;.
\end{eqnarray}
The field $\tilde{\psi}$ decouples from the Lagrangian and we shall not
consider it further.

The free Lagrangian density of the graviton is obtained by simplifying
the first two terms above
\begin{equation}
L_{0g}=-\frac{1}{2}\left\{\partial_{\lambda}h^{\mu\nu}\partial^{\lambda}
h_{\mu\nu}+\partial^{\lambda}h^{\mu}_{\;\;\mu}
\partial_{\lambda}h^{\nu}_{\;\;\nu}
-2\partial_{\mu}h^{\mu\nu}\partial_{\nu}h^{\lambda}_{\;\;\lambda}\right\}
+\partial^{\nu}h_{\mu\nu}B^{\mu}+\frac{1}{4}B_{\mu}B^{\mu}
\end{equation}
where we added gauge fixing terms in the form of a Lagrange multiplier
field $B_{\mu}$. 

Eliminating $B_{\mu}$ from its field equation leaves us with the
Lagrangian
\begin{equation}\label{Lag4}
\tilde{L}_{0g}=-\frac{1}{2}\left\{\partial_{\lambda}h^{\mu\nu}\partial^{\lambda}h_{\mu\nu}+\partial^{\lambda}h^{\mu}_{\;\;\mu}
\partial_{\lambda}h^{\nu}_{\;\;\nu}
-2\partial_{\mu}h^{\mu\nu}\partial_{\nu}h^{\lambda}_{\;\;\lambda}\right\}
-\partial^{\nu}h_{\mu\nu}\partial_{\lambda}h^{\mu\lambda}\;\;.
\end{equation}
whose  canonical quantization we shall now undertake. 


Temporarily setting the scalar fields to zero, we obtain
\begin{equation}\label{feq1a}
\Box h_{\mu\nu}+\eta_{\mu\nu}\Box h^{\lambda}_{\;\;\lambda}
-\eta_{\mu\nu}\partial_{\lambda}\partial_{\rho}h^{\lambda\rho}
-\partial_{\mu}\partial_{\nu}h^{\lambda}_{\;\;\lambda}
+\partial_{\mu}\partial^{\lambda}h_{\nu\lambda}
+\partial_{\nu}\partial^{\lambda}h_{\mu\lambda}=0\;\;.
\end{equation}
for the graviton field equation. Its trace is
\begin{equation}
\Box h^{\lambda}_{\;\;\lambda}=
\frac{n-2}{n}\partial_{\lambda}\partial_{\rho} h^{\lambda\rho}
\end{equation}
implying that (\ref{feq1a}) becomes
\begin{equation}\label{feq2a}
\Box h_{\mu\nu}
-\frac{2}{n}\eta_{\mu\nu}\partial_{\lambda}\partial_{\rho}h^{\lambda\rho}
-\partial_{\mu}\partial_{\nu}h^{\lambda}_{\;\;\lambda}
+\partial_{\mu}\partial^{\lambda}h_{\nu\lambda}
+\partial_{\nu}\partial^{\lambda}h_{\mu\lambda}=0\;\;.
\end{equation}
Taking  the $\partial^{\nu}$ derivative of (\ref{feq2a}) leads to
\begin{equation}\label{feq3a}
\Box\partial^{\nu}h_{\mu\nu}=0\;\; .
\end{equation}
This, along with the D'Alembertians of (\ref{feq2a}) 
and (\ref{feq3a}) respectively lead to
\begin{equation}\label{feqAA}
\Box^{2}h_{\mu\nu}-\partial_{\mu}\partial_{\nu}\Box h^{\lambda}_{\;\;\lambda}
=0\;\; \quad \mbox{and} \quad   \Box^{2}h^{\lambda}_{\;\;\lambda}=0\;\;
\end{equation}
which finally implies
\begin{equation}\label{feq4a}
\Box^{3}h_{\mu\nu}=0
\end{equation}
a relation characteristic of $n\geq 3$ Lagrangians.

The conjugate momentum is
\begin{equation}\label{mom}
\pi^{\mu\nu}=\partial_{0}h^{\mu\nu}+\eta^{\mu\nu}
\left(\partial_{0}h^{\lambda}_{\;\;\lambda}
+\partial_{\lambda}h^{\lambda}_{\;\;0}\right)
+\eta^{\mu 0}\left(\frac{1}{2}\partial^{\nu}h^{\lambda}_{\;\;\lambda}
-\partial_{\lambda}h^{\nu\lambda}\right)
+\eta^{\nu 0}\left(\frac{1}{2}\partial^{\mu}h^{\lambda}_{\;\;\lambda}
-\partial_{\lambda}h^{\mu\lambda}\right)\;\;.
\end{equation}
which implies 
\begin{eqnarray}\label{comp2}
\partial_{0}h_{00}&=&\frac{1}{2}\pi^{00}+\frac{1}{2}\partial_{i}h_{0i}
\nonumber \\
\partial_{0}h_{0i}&=&-\frac{1}{2}\pi^{0i}+\frac{1}{4}\partial_{i}h_{00}
-\frac{1}{4}\partial_{i}h_{jj}+\frac{1}{2}\partial_{j}h_{ij}
\\
\partial_{0}h_{ij}&=&\pi^{ij}-\frac{\delta_{ij}}{n}\pi^{kk}
+\frac{\delta_{ij}}{n}\partial_{k}h_{0k}\;\;.
\nonumber
\end{eqnarray}
The equal-time commutation relations are 
\begin{eqnarray}
\left[h_{\mu\nu}(x), \;\pi^{\lambda\rho}(y)\right]_{eq}
&=&\frac{i}{2}\left(\delta^{\lambda}_{\mu}\delta^{\rho}_{\nu}
+\delta^{\rho}_{\mu}\delta^{\lambda}_{\nu}\right)\delta^{(n-1)}(x-y)
\nonumber \\
\\
\left[h_{\mu\nu}(x), \;h_{\lambda\rho}(y)\right]_{eq}
&=&\left[\pi^{\mu\nu}(x), \;\pi^{\lambda\rho}(y)\right]_{eq}=0\;\;.
\nonumber 
\end{eqnarray}
implying that the commutators between $h_{\mu\nu}$ and 
$\partial_{0}h_{\lambda\rho}$ become
\begin{eqnarray}\label{commu1}
&&\left[h_{\mu\nu},\;\partial_{0}h_{\kappa\sigma}\right]_{eq} 
\nonumber \\
&&=\frac{i}{2}\left\{\frac{1}{2}\left(\eta_{\mu\kappa}\eta_{\nu\sigma}
+\eta_{\mu\sigma}\eta_{\nu\kappa}\right)
+\frac{1}{2}\left(\bar{\eta}_{\mu\kappa}\bar{\eta}_{\nu\sigma}
+\bar{\eta}_{\mu\sigma}\bar{\eta}_{\nu\kappa}\right)
-\frac{2}{n}\bar{\eta}_{\mu\nu}\bar{\eta}_{\kappa\sigma}\right\}
\delta^{(n-1)}(x-y)
\end{eqnarray}
where
\begin{equation}
\bar{\eta}_{\mu\nu}\equiv \eta_{\mu\nu}+\eta_{\mu 0}\eta_{\nu 0}\;\;.
\end{equation}
The proof of this relation is given in the Appendix.

The solution to (\ref{feq4a}) is
\begin{eqnarray}\label{h}
h_{\mu\nu}(x)&=&-\int d^{n-1}z\;D^{(n)}(x-z)\bar{\partial}^{z}_{0}h_{\mu\nu}(z)
-\int d^{n-1}z\;\tilde{D}^{(n)}(x-z)\bar{\partial}^{z}_{0}\Box h_{\mu\nu}(z)
\nonumber \\
&&-\int d^{n-1}z\;\tilde{\tilde{D}}^{(n)}(x-z)\bar{\partial}^{z}_{0}\Box^{2}
h_{\mu\nu}(z)
\;\;.
\end{eqnarray}
where  $D^{(n)}$, $\tilde{D}^{(n)}$ and $\tilde{\tilde{D}}^{(n)}$ are 
defined via
\begin{eqnarray}
D^{(n)}(x)
&=&-\frac{i}{(2\pi)^{n-1}}\int d^{n}k\;\epsilon(k_{0})\delta(k^{2})e^{ikx}
\nonumber \\
\tilde{D}^{(n)}(x)
&=&-\frac{i}{(2\pi)^{n-1}}\int d^{n}k\;\epsilon(k_{0})\delta^{\prime}
(k^{2})e^{ikx}
\nonumber \\  
\tilde{\tilde{D}}^{(n)}(x)
&=&-\frac{i}{(2\pi)^{n-1}}\int d^{n}k\;\epsilon(k_{0})\delta^{\prime\prime}
(k^{2})e^{ikx}
\nonumber
\end{eqnarray}
We next need to express all of $h_{\mu\nu}, \partial_{0}h_{\mu\nu},
\Box \partial_{0}h_{\mu\nu}, \Box^{2}h_{\mu\nu}$ and 
$\Box^{2}\partial_{0}h_{\mu\nu}$ in terms of the canonical variables
and calculate commutators at equal-time.  This rather lengthy and
complicated calculation is given in the Appendix.

{}From (\ref{h}) and the equal-time commutators, 
the commutator among the components of $h_{\mu\nu}$ at two 
arbitrary space-time points can be calculated :
\begin{eqnarray}\label{commu2}
\lefteqn{\left[h_{\mu\nu}(x),\;h_{\lambda\rho}(y)\right]}
\nonumber \\
&&=\frac{i}{2}\left(\eta_{\mu\lambda}\eta_{\nu\rho}
+\eta_{\mu\rho}\eta_{\nu\lambda}
-\frac{2}{n}\eta_{\mu\nu}\eta_{\lambda\rho}\right)D^{(n)}(x-y)
\nonumber \\
&&+\frac{i}{4}\left\{-\eta_{\mu\lambda}\partial_{\nu}\partial_{\rho}
-\eta_{\mu\rho}\partial_{\nu}\partial_{\lambda}
-\eta_{\nu\lambda}\partial_{\mu}\partial_{\rho}
-\eta_{\nu\rho}\partial_{\mu}\partial_{\lambda}
+\frac{4}{n}\left(\eta_{\mu\nu}\partial_{\lambda}\partial_{\rho}
+\eta_{\lambda\rho}\partial_{\mu}\partial_{\nu}\right)\right\}
\tilde{D}^{(n)}(x-y)
\nonumber \\
&&+\frac{i}{2}\left(1-\frac{2}{n}\right)\partial_{\mu}\partial_{\nu}
\partial_{\lambda}\partial_{\rho}\tilde{\tilde{D}}^{(n)}(x-y)\;\;.
\end{eqnarray}
This expression is valid even when $n=2$. 
The graviton propagator is 
\begin{equation}
\langle 0|T(h_{\mu\nu}(x)h_{\lambda\rho}(y))|0\rangle 
=-\frac{i}{2(2\pi)^{n}}
\int d^{n}k\;e^{ik(x-y)}\frac{X_{\mu\nu,\lambda\rho}}{k^{2}-i\epsilon}
\end{equation}
where
\begin{eqnarray}\label{propag}
X_{\mu\nu,\lambda\rho}&=&\eta_{\mu\lambda}\eta_{\nu\rho}
+\eta_{\mu\rho}\eta_{\nu\lambda}-\frac{2}{n}\eta_{\mu\nu}\eta_{\lambda\rho}
\nonumber \\
&&+\frac{1}{2k^{2}}\left\{-\eta_{\mu\lambda}k_{\nu}k_{\rho}
-\eta_{\mu\rho}k_{\nu}k_{\lambda}-\eta_{\nu\lambda}k_{\mu}k_{\rho}
-\eta_{\nu\rho}k_{\mu}k_{\lambda}
+\frac{4}{n}\left(\eta_{\mu\nu}k_{\lambda}k_{\rho}
+\eta_{\lambda\rho}k_{\mu}k_{\nu}\right)\right\}
\nonumber \\
&&+\left(1-\frac{2}{n}\right)\frac{k_{\mu}k_{\nu}k_{\lambda}k_{\rho}}
{(k^{2})^{2}}\;\;.
\end{eqnarray}

We turn now to the scalar fields, whose Lagrangian density to 
lowest order in the graviton coupling is
\begin{equation}
L_{int}=-\frac{1}{2}\left\{\frac{1}{2}\eta^{\mu\nu}
\left(\phi^{,\alpha}\phi_{,\alpha}+m^{2}\phi^{2}\right)
-\phi^{,\mu}\phi^{,\nu}\right\}h_{\mu\nu}\;\;.
\end{equation}
The one graviton exchange diagram yields the $S$-matrix element 
\begin{eqnarray}
S&=&\frac{4\pi iG_{n}}{(2\pi)^{n-2}}
\left(p^{0}_{1}p^{0}_{2}q^{0}_{1}q^{0}_{2}\right)^{-1/2}
\left[p^{\mu}_{1}q^{\nu}_{1}-\frac{1}{2}\eta^{\mu\nu}
(p_{1}\cdot q_{1}+m^{2}_{1})\right]
\left[p^{\lambda}_{2}q^{\rho}_{2}-\frac{1}{2}\eta^{\lambda\rho}
(p_{2}\cdot q_{2}+m^{2}_{2})\right]
\nonumber \\
&&\makebox[5em]{}\times \frac{X_{\mu\nu,\lambda\rho}}{k^{2}}
\delta^{(n)}(p_{1}+p_{2}-q_{1}-q_{2})
\nonumber 
\end{eqnarray}
where $p^{\mu}_{a}$, $q^{\mu}_{a}$ and $k^{\mu}$ are the four-
momenta of the initial particles, the final particles and the 
transferred graviton, respectively. This result is valid for $n=2$ also.
In the lowest order and the static approximation $T$-matrix element 
is
\begin{equation}
T_{n}=-4\left(1-\frac{1}{n}\right)\frac{G_{n}}{(2\pi)^{n-2}}
\frac{m_{1}m_{2}}{{\bf k}^{2}}\;\;.
\end{equation}
whose associated potential is $V = \int d^n k e^{-ikx}T(k)$
in $n$ dimensions. 

The $T$-matrix elements for $n=2, 3$ and $4$ are
\begin{eqnarray}
T_{2}&=&-\frac{2G_{2}m_{1}m_{2}}{{\bf k^{2}}}
\\
T_{3}&=&-\frac{8}{3}\cdot\frac{G_{3}}{(2\pi)}\cdot
\frac{m_{1}m_{2}}{{\bf k^{2}}}
\\
T_{4}&=&-\frac{3G_{4}}{(2\pi)^{2}}\cdot\frac{m_{1}m_{2}}{{\bf k^{2}}}\;\;.
\end{eqnarray}
and the corresponding potentials are
\begin{eqnarray}
V_{2}&=&2\pi G_{2}m_{1}m_{2}\;r \label{pot2}
\\
V_{3}&=&2\left(\frac{4}{3}G_{3}\right)m_{1}m_{2}\mbox{log}\;r \label{pot3}
\\
V_{4}&=&-\frac{3}{2}\cdot\frac{G_{4}m_{1}m_{2}}{r} \label{pot4}
\end{eqnarray}
By identifying the  gravitational constants as
\begin{equation}
G_{N,2}\equiv G_{2} \qquad G_{N,3}\equiv \frac{4}{3}G_{3}
\qquad G_{N,4}\equiv \frac{3}{2}G_{4}
\end{equation}
we get the correct Newtonian potentials in each dimension.

The above results are in strong contrast with $(d+1)$ dimensional
general relativity, whose free Lagrangian density is
\begin{eqnarray}
L_{0g}&=&-\frac{1}{2}\left\{\partial_{\lambda}h^{\mu\nu}\partial^{\lambda}
h_{\mu\nu}-\partial^{\lambda}h^{\mu}_{\;\;\mu}
\partial_{\lambda}h^{\nu}_{\;\;\nu}
-2\partial_{\mu}h^{\mu\nu}\partial^{\lambda}h_{\lambda\nu}
+2\partial_{\mu}h^{\mu\nu}\partial_{\nu}h^{\lambda}_{\;\;\lambda}\right\}
\nonumber \\
&&+\left(\partial^{\nu}h_{\nu\mu}-\frac{1}{2}\partial_{\mu}h^{\lambda}
_{\;\;\lambda}\right)B^{\mu}+\frac{1}{4}B_{\mu}B^{\mu}
\nonumber 
\end{eqnarray}
where gauge fixing terms have been added. A computation analogous to
the one above gives
\[
\langle 0|T(h_{\mu\nu}(x)h_{\lambda\rho}(y)|0\rangle =-\frac{i}{2(2\pi)^{n}}
\int d^{n}k\;e^{ik(x-y)}\frac{X_{\mu\nu,\lambda\rho}}{k^{2}-i\epsilon}
\]
where
\[
X_{\mu\nu,\lambda\rho}=\eta_{\mu\lambda}\eta_{\nu\rho}
+\eta_{\mu\rho}\eta_{\nu\lambda}-\frac{2}{n-2}\eta_{\mu\nu}\eta_{\lambda\rho}
\]
The $S$-matrix element of one graviton exchange diagram is
\begin{eqnarray}
S&=&\frac{4\pi iG_{n}}{(2\pi)^{n-2}}
\left(p^{0}_{1}p^{0}_{2}q^{0}_{1}q^{0}_{2}\right)^{-1/2}
\left[p^{\mu}_{1}q^{\nu}_{1}-\frac{1}{2}\eta^{\mu\nu}
(p_{1}\cdot q_{1}+m^{2}_{1})\right]
\left[p^{\alpha}_{2}q^{\beta}_{2}-\frac{1}{2}\eta^{\alpha\beta}
(p_{2}\cdot q_{2}+m^{2}_{2})\right]
\nonumber \\
&&\makebox[5em]{}\times \frac{X_{\mu\nu,\alpha\beta}}{k^{2}}
\delta^{(n)}(p_{1}+p_{2}-q_{1}-q_{2})
\nonumber 
\end{eqnarray}
which in turn yields the $T$-matrix element 
\begin{equation}
T=-\frac{4G_{n}}{(2\pi)^{n-2}}\cdot \frac{n-3}{n-2}\cdot 
\frac{m_{1}m_{2}}{k^{2}}
\end{equation}
in the static appoximation in $n$ dimensions.

The potential for $n=4$ is
\begin{equation}
V^{(4)}=-\frac{G_{4}m_{1}m_{2}}{r}
\end{equation}
in agreement with (\ref{pot4}). However the potential for $n=3$
vanishes, and the potential for $n=2$ diverges.  This latter
situation can be dealt with by setting $G_{n}=(1-\frac{n}{2})G_{2}$ 
and taking the $n\rightarrow 2$ limit \cite{2dross},
which yields the 2-dimensional $T$-matix element
\begin{equation}
T=-\frac{2G_{2}m_{1}m_{2}}{k^{2}}
\end{equation}
and potential
\begin{equation}
V^{(2)}=2\pi G_{2}m_{1}m_{2}r
\end{equation}

Unlike general relativity, 3-dimensional dilaton gravity includes the 
Newtonian potential in any dimension, once the gravitational constant
is appropriately rescaled. In this sense the theory of
gravity (\ref{act1}) we consider is a relativistic extension of 
Newtonian gravity in $(d+1)$ dimensions. General relativity, on the
other hand, does not include Newtonian gravity in $(2+1)$ dimensions
and is empty in $(1+1)$ dimensions. In the latter case an appropriate
rescaling of Newton's constant yields the theory (\ref{act1}) in
the $n\to 2$ limit \cite{2dross}.

\section{Conclusions}

We have obtained an exact self-consistent solution to the 2-body problem
in a $(1+1)$ dimensional theory of gravity with a Newtonian limit.
To our knowledge, this is the only exact relativistic 2-body solution
of this type. We are able to explore all possible limits of this solution,
including large and small gravitational coupling and/or mass and/or momenta.

A natural extension of what we have done would be to attempt to
solve the $N$ body problem. It would also be of interest to couple
other matter fields (e.g. electromagnetism), and to investigate the
extent to which our methods are applicable to other dilaton theories
of gravity.  

Finally, and what is perhaps most interesting, is to quantize the degrees
of freedom of the 2 body system we consider based on the Hamiltonian
given in (\ref{eqH1}).  The quantum theory based on (ref{eqH1}) is
a quantum theory of gravity coupled to matter whose slow-motion
weak field limits should be straightforwardly comparble to 
that of the non-relativistic mechanics of 2 particles in a linear
confining potential.  As such it should offer interesting insights
into the behaviour of quantum gravity.

\section*{Acknowledgements} This work was supported in part by the
Natural Sciences and Engineering Research Council of Canada.

\section*{Appendix: Commutation relations}          

We use the notation
\[
\bar{\eta}_{\mu\nu}=\eta_{\mu\nu}+\eta_{\mu 0}\eta_{\nu 0}
\qquad \bar{\partial}_{\mu}=\partial_{\mu}+\eta_{\mu 0}\partial_{0}
\quad .
\]
The general form of the equal time commutation relations is
\begin{eqnarray}\label{gencom}
\lefteqn{\left[h_{\mu\nu}(x), \;\pi^{\lambda\rho}(y)\right]_{eq}}
\nonumber \\
&=&\left[h_{\mu\nu}(x), \;\partial_{0}h^{\mu\nu}+\eta^{\mu\nu}
\left(\partial_{0}h^{\lambda}_{\;\;\lambda}
+\partial_{\lambda}h^{\lambda 0}\right)
+\eta^{\mu 0}\left(\frac{1}{2}\partial^{\nu}h^{\lambda}_{\;\;\lambda}
-\partial_{\lambda}h^{\nu\lambda}\right)
+\eta^{\nu 0}\left(\frac{1}{2}\partial^{\mu}h^{\lambda}_{\;\;\lambda}
-\partial_{\lambda}h^{\mu\lambda}\right)\right]_{eq}
\nonumber \\
&=&\frac{i}{2}\left(\delta^{\lambda}_{\mu}\delta^{\rho}_{\nu}
+\delta^{\rho}_{\mu}\delta^{\lambda}_{\nu}\right)\delta^{(n-1)}(x-y)
\end{eqnarray}
from which we shall now deduce various commutators of interest.

Taking the trace of (\ref{gencom}) implies the relations
\begin{eqnarray}
&&\left[h_{\mu\nu},\;n\partial_{0}h^{\alpha}_{\;\;\alpha}
-(n-2)\partial_{\alpha}h_{0}^{\;\;\alpha}\right]_{eq}=
i\eta_{\mu\nu}\delta^{(n-1)}(x-y)
\nonumber \\
&&\left[h_{\mu\nu},\;\partial_{0}h^{\alpha}_{\;\;\alpha}
-\partial_{\alpha}h_{0}^{\;\;\alpha}\right]_{eq}
=\frac{2}{n}\left[h_{\mu\nu},\;\partial_{0}h_{00}\right]_{eq}
+\frac{i}{n}\eta_{\mu\nu}\delta^{(n-1)}(x-y)
\\
&&\left[h_{\mu\nu},\;\partial_{0}h^{\alpha}_{\;\;\alpha}\right]_{eq}
=\frac{2-n}{n}\left[h_{\mu\nu},\;\partial_{0}h_{00}\right]_{eq}
+\frac{i}{n}\eta_{\mu\nu}\delta^{(n-1)}(x-y)
\nonumber
\end{eqnarray}
from which it follows that
\begin{eqnarray}
&&\left[h_{\mu\nu},\;\partial_{0}h_{\kappa\sigma}
+\left(\frac{2}{n}\bar{\eta}_{\kappa\sigma}-\eta_{\kappa 0}\eta_{\sigma 0}
\right)\partial_{0}h_{00}-\eta_{\kappa 0}\partial_{0}h_{0 \sigma}
-\eta_{\sigma 0}\partial_{0}h_{0 \kappa}\right]_{eq}
\nonumber \\
&&=\frac{i}{2}\left(\eta_{\mu\kappa}\eta_{\nu\sigma}
+\eta_{\mu\sigma}\eta_{\nu\kappa}
-\frac{2}{n}\eta_{\mu\nu}\bar{\eta}_{\kappa\sigma}\right)
\delta^{(n-1)}(x-y)
\end{eqnarray}
The  $(\kappa, \sigma)=(0, \sigma)$ component gives
\begin{equation}
\left[h_{\mu\nu},\;\partial_{0}h_{0 \sigma}\right]_{eq}
=\frac{i}{4}\left(\eta_{\mu 0}\eta_{\nu\sigma}
+\eta_{\mu\sigma}\eta_{\nu 0}\right)\delta^{(n-1)}(x-y)
\end{equation}
which yields in turn the commutation relation
\begin{eqnarray}
&&\left[h_{\mu\nu},\;\partial_{0}h_{\kappa\sigma}\right]_{eq} 
\nonumber \\
&&=\frac{i}{2}\left\{\frac{1}{2}\left(\eta_{\mu\kappa}\eta_{\nu\sigma}
+\eta_{\mu\sigma}\eta_{\nu\kappa}\right)
+\frac{1}{2}\left(\bar{\eta}_{\mu\kappa}\bar{\eta}_{\nu\sigma}
+\bar{\eta}_{\mu\sigma}\bar{\eta}_{\nu\kappa}\right)
-\frac{2}{n}\bar{\eta}_{\mu\nu}\bar{\eta}_{\kappa\sigma}\right\}
\delta^{(n-1)}(x-y)
\end{eqnarray}
We now consider the problem of expressing $\Box h_{\mu\nu}, 
\Box\partial_{0}h_{\mu\nu}, \Box^{2}h_{\mu\nu}$ and 
$\Box^{2}\partial_{0}h_{\mu\nu}$ in terms of $h_{\mu\nu}$ and 
$\partial_{0}h_{\mu\nu}$ (or equivalently the canonical variables).
The components of the first-order field equations (\ref{feq1a})
give
\begin{eqnarray}\label{feqA1}
\Box h_{00}&=&\frac{1}{2}\triangle h_{ii}-\frac{1}{2}\partial_{i}\partial_{j}
h_{ij}
\nonumber \\
\Box h_{0i}&=&\frac{1}{2}\partial_{i}\partial_{0}h_{jj}
-\frac{1}{2}\partial_{j}\partial_{0}h_{ij}-\frac{1}{2}\partial_{i}\partial_{j}
h_{0j}+\frac{1}{2}\triangle h_{0i}
\nonumber \\
\\
\Box h_{ij}&=&-\frac{4}{n}\delta_{ij}\partial_{k}\partial_{0}h_{0k}
+\partial_{i}\partial_{0}h_{j0}+\partial_{j}\partial_{0}h_{i0}
\nonumber \\
&&+\frac{1}{n}\delta_{ij}\left(3\partial_{k}\partial_{l}h_{kl}
+2\triangle h_{00}-\triangle h_{kk}\right)
\nonumber \\
&&-\partial_{i}\partial_{j}h_{00}+\partial_{i}\partial_{j}h_{kk}
-\partial_{i}\partial_{k}h_{jk}-\partial_{j}\partial_{k}h_{ik}
\nonumber 
\end{eqnarray}
and
\begin{eqnarray}\label{feqA2}
\partial^{2}_{0}h_{00}&=&\frac{1}{2}\partial_{i}\partial_{j}h_{ij}
+\triangle h_{00}-\frac{1}{2}\triangle h_{ii}
\nonumber \\
\partial^{2}_{0}h_{0i}&=&-\frac{1}{2}\partial_{i}\partial_{0}h_{jj}
+\frac{1}{2}\partial_{j}\partial_{0}h_{ij}+\partial_{i}\partial_{j}h_{0j}
+\frac{1}{2}\triangle h_{0i}
\nonumber \\
\\
\partial^{2}_{0}h_{ij}&=&\frac{4}{n}\delta_{ij}\partial_{k}\partial_{0}h_{0k}
-\partial_{i}\partial_{0}h_{j0}-\partial_{j}\partial_{0}h_{i0}
\nonumber \\
&&-\frac{1}{n}\delta_{ij}\left(3\partial_{k}\partial_{l}h_{kl}
+2\triangle h_{00}-\triangle h_{kk}\right)
\nonumber \\
&&+\partial_{i}\partial_{j}h_{00}-\partial_{i}\partial_{j}h_{kk}
+\partial_{i}\partial_{k}h_{jk}+\partial_{j}\partial_{k}h_{ik}
+\triangle h_{ij}
\nonumber
\end{eqnarray}
{}From (\ref{feqA1}) and (\ref{feqA2}) we get 
\begin{eqnarray}\label{feqA3}
\Box\partial_{0}h_{00}&=&\frac{1}{2}\triangle\partial_{0}h_{ii}
-\frac{1}{2}\partial_{i}\partial_{j}\partial_{0}h_{ij}
\nonumber \\
\Box\partial_{0}h_{0i}&=&\triangle\partial_{0}h_{0i}
+\left(1-\frac{4}{n}\right)\partial_{i}\partial_{j}\partial_{0}h_{0j}
+\left(\frac{2}{n}-1\right)\triangle\partial_{i}h_{00}
+\left(1-\frac{1}{n}\right)\triangle\partial_{i}h_{jj}
\nonumber \\
&&-\triangle\partial_{j}h_{ij}+\left(\frac{3}{n}-1\right)
\partial_{i}\partial_{j}\partial_{k}h_{jk}
\nonumber \\
\\
\Box\partial_{0}h_{ij}&=&\frac{1}{n}\delta_{ij}\left(\triangle\partial_{0}h_{kk}+\partial_{k}\partial_{l}\partial_{0}h_{kl}+2\triangle\partial_{0}h_{00}
-4\triangle\partial_{k}h_{0k}\right)
\nonumber \\
&&-\partial_{i}\partial_{j}\partial_{0}h_{00}
-\frac{1}{2}\partial_{i}\partial_{k}\partial_{0}h_{jk}
-\frac{1}{2}\partial_{j}\partial_{k}\partial_{0}h_{ik}
+\partial_{i}\partial_{j}\partial_{k}h_{0k}
\nonumber \\
&&+\frac{1}{2}\triangle\partial_{i}h_{0j}+\frac{1}{2}\triangle\partial_{j}h_{0i}\nonumber 
\end{eqnarray}
whereas (\ref{feqAA}) implies
\begin{eqnarray}\label{feqA4}
\Box^{2}h_{00}&=&\left(1-\frac{2}{n}\right)\left\{
-2\triangle\partial_{i}\partial_{0}h_{0i}+\triangle^{2}h_{00}
-\frac{1}{2}\triangle^{2}h_{ii}
+\frac{3}{2}\triangle\partial_{i}\partial_{j}h_{ij}\right\}
\nonumber \\
\Box^{2}h_{0i}&=&\left(1-\frac{2}{n}\right)\partial_{i}\left\{
\triangle\partial_{0}h_{00}+\frac{1}{2}\triangle\partial_{0}h_{jj}
-2\triangle\partial_{j}h_{0j}
+\frac{1}{2}\partial_{j}\partial_{k}\partial_{0}h_{jk}\right\}
\\
\Box^{2}h_{ij}&=&\left(1-\frac{2}{n}\right)\partial_{i}\partial_{j}\left\{
-2\partial_{k}\partial_{0}h_{0k}+\triangle h_{00}
-\frac{1}{2}\triangle h_{kk}+\frac{3}{2}\partial_{k}\partial_{l}h_{kl}
\right\}
\nonumber 
\end{eqnarray}
and
\begin{eqnarray}\label{feqA5}
\Box^{2}\partial_{0}h_{00}&=&\left(1-\frac{2}{n}\right)\left\{
\triangle^{2}\partial_{0}h_{00}+\frac{1}{2}\triangle^{2}\partial_{0}h_{ii}
+\frac{1}{2}\partial_{i}\partial_{j}\partial_{0}h_{ij}
-2\triangle^{2}\partial_{i}h_{0i}\right\}
\nonumber \\
\Box^{2}\partial_{0}h_{0i}&=&\left(1-\frac{2}{n}\right)\partial_{i}
\left\{-2\triangle\partial_{j}\partial_{0}h_{0j}+\triangle^{2}h_{00}
-\frac{1}{2}\triangle^{2}h_{jj}
+\frac{3}{2}\triangle\partial_{j}\partial_{k}h_{jk}\right\}
\\
\Box^{2}\partial_{0}h_{ij}&=&\left(1-\frac{2}{n}\right)\partial_{i}\partial_{j}
\left\{\triangle\partial_{0}h_{00}+\frac{1}{2}\triangle\partial_{0}h_{kk}
+\frac{1}{2}\partial_{k}\partial_{l}\partial_{0}h_{kl}
-2\triangle\partial_{k}h_{0k}\right\}
\nonumber 
\end{eqnarray}

To calculate the commutator $\left[h_{\mu\nu}(x),\;h_{\lambda\rho}(y)
\right]$ at two arbitrary space-time points, we first express 
$h_{\lambda\rho}(y)$ as
\begin{eqnarray}
h_{\lambda\rho}(y)&=&-\int d^{n-1}z\;D^{(n)}(y-z)\bar{\partial}^{z}_{0}
h_{\lambda\rho}(z)
-\int d^{n-1}z\;\tilde{D}^{(n)}(y-z)\bar{\partial}^{z}_{0}
\Box h_{\lambda\rho}(z)
\nonumber \\
&&-\int d^{n-1}z\;\tilde{\tilde{D}}^{(n)}(y-z)\bar{\partial}^{z}_{0}
\Box^{2}h_{\lambda\rho}(z)
\;\;.
\end{eqnarray}
\begin{eqnarray}
\lefteqn{\left[h_{\mu\nu}(x),\;h_{\lambda\rho}(y)\right]}
\nonumber \\
&=&-\int d^{n-1}z: D^{(n)}(y-z)\left[h_{\mu\nu}(x),\;\partial_{0}
h_{\lambda\rho}(z)\right]_{eq}
-\partial_{0}D^{(n)}(y-z)\left[h_{\mu\nu}(x),\;h_{\lambda\rho}(z)\right]_{eq}
\nonumber \\
&&\makebox[5em]{}+\tilde{D}^{(n)}(y-z)\left[h_{\mu\nu}(x),\;
\partial_{0}\Box h_{\lambda\rho}(z)\right]_{eq}
-\partial_{0}\tilde{D}^{(n)}(y-z)\left[h_{\mu\nu}(x),\;
\Box h_{\lambda\rho}\right]_{eq}
\nonumber \\
&&\makebox[5em]{}+\tilde{\tilde{D}}^{(n)}(y-z)\left[h_{\mu\nu}(x),\;
\partial_{0}\Box^{2}h_{\lambda\rho}(z)\right]_{eq}
-\partial_{0}\tilde{\tilde{D}}^{(n)}(y-z)\left[h_{\mu\nu},\;
\Box^{2}h_{\lambda\rho}(z)\right]_{eq}
\nonumber \\
\end{eqnarray}
We shall refer to the 6 terms on the RHS as
term 1, term 2 $\cdot\cdot\cdot$ term 6 respectively. From 
the commutator (\ref{commu1}), terms 1 and 2 are
\begin{eqnarray}
\langle \mbox{term 1}\rangle 
&=& \frac{i}{4}\left\{\eta_{\mu\lambda}\eta_{\nu\rho}
+\eta_{\mu\rho}\eta_{\nu\lambda}+\bar{\eta}_{\mu\lambda}\bar{\eta}_{\nu\rho}
+\bar{\eta}_{\mu\rho}\bar{\eta}_{\nu\lambda}-\frac{4}{n}\bar{\eta}_{\mu\nu}
\bar{\eta}_{\lambda\rho}\right\}D^{(n)}(x-y)\label{term1}
\nonumber \\
\langle \mbox{term 2}\rangle &=&0 \;\;.\label{term2}
\end{eqnarray}
For terms 3 and 4 we shall calculate the equal-time commutators 
$\left[h_{\mu\nu}(x),\;\partial_{0}\Box h_{\lambda\rho}\right]_{eq}$
and $\left[h_{\mu\nu}(x),\;\Box h_{\lambda\rho}(z)\right]_{eq}$ 
by using the expressions (\ref{feqA1}) and (\ref{feqA3}). 
Putting terms 3 and 4 together gives
\begin{eqnarray}\label{term34}
\lefteqn{\langle \mbox{term 3}+\mbox{term 4}\rangle}
\nonumber \\
&=&\frac{i}{4}\left\{-\bar{\eta}_{\mu\lambda}\partial_{\nu}\partial_{\rho}
-\bar{\eta}_{\mu\rho}\partial_{\nu}\partial_{\lambda}
-\bar{\eta}_{\nu\lambda}\partial_{\mu}\partial_{\rho}
-\bar{\eta}_{\nu\rho}\partial_{\mu}\partial_{\lambda}\right.
\nonumber \\
&&\left.+\frac{4}{n}\left(\bar{\eta}_{\mu\nu}\partial_{\lambda}\partial_{\rho}
+\bar{\eta}_{\lambda\rho}\partial_{\mu}\partial_{\nu}\right)
-2\eta_{\mu 0}\eta_{\nu 0}\bar{\partial}_{\lambda}\bar{\partial}_{\rho}
-2\eta_{\lambda 0}\eta_{\rho 0}\bar{\partial}_{\mu}\bar{\partial}_{\nu}\right.
\nonumber \\
&&\left.-\left(1-\frac{4}{n}\right)\left(\eta_{\mu 0}\eta_{\lambda 0}
\bar{\partial}_{\nu}\bar{\partial}_{\rho}
+\eta_{\mu 0}\eta_{\rho 0}\bar{\partial}_{\nu}\bar{\partial}_{\lambda}
+\eta_{\nu 0}\eta_{\lambda 0}\bar{\partial}_{\mu}\bar{\partial}_{\rho}
+\eta_{\nu 0}\eta_{\rho 0}\bar{\partial}_{\mu}\bar{\partial}_{\lambda}
\right)\right\}\tilde{D}^{(n)}(x-y)
\nonumber \\
&&+\frac{i}{4}\left\{\frac{4}{n}\left(\eta_{\mu 0}\eta_{\nu 0}
\bar{\eta}_{\lambda\rho}+\eta_{\lambda 0}\eta_{\rho 0}\bar{\eta}_{\mu\nu}
\right)\right.
\nonumber \\
&&\left.-\left(\eta_{\mu 0}\eta_{\lambda 0}\bar{\eta}_{\nu\rho}
+\eta_{\mu 0}\eta_{\rho 0}\bar{\eta}_{\nu\lambda}
+\eta_{\nu 0}\eta_{\lambda 0}\bar{\eta}_{\mu\rho}
+\eta_{\nu 0}\eta_{\rho 0}\bar{\eta}_{\mu\lambda}\right)\right\}D^{(n)}(x-y)
\end{eqnarray}
where we used the relation
\[
\triangle\tilde{D}^{(n)}=\partial^{2}_{0}\tilde{D}^{(n)}+D^{(n)}\;\;.
\]

Similarly by calculating the equal-time commutators 
$\left[h_{\mu\nu},\;\partial_{0}\Box^{2}h_{\lambda\rho}\right]_{eq}$
and $\left[h_{\mu\nu}(x),\;\Box^{2}h_{\lambda\rho}(z)\right]_{eq}$
(making use of the expressions (\ref{feqA4}) and (\ref{feqA5})), we have
\begin{eqnarray}\label{term56}
\lefteqn{\langle \mbox{term 5}+\mbox{term 6}\rangle}
\nonumber \\
&=&\frac{i}{2}\left(1-\frac{2}{n}\right)\partial_{\mu}\partial_{\nu}
\partial_{\lambda}\partial_{\rho}\tilde{\tilde{D}}^{(n)}(x-y)
\nonumber \\
&&+\frac{i}{2}\left(1-\frac{2}{n}\right)\left\{
\eta_{\mu 0}\eta_{\lambda 0}\partial_{\nu}\partial_{\rho}
+\eta_{\mu 0}\eta_{\rho 0}\partial_{\nu}\partial_{\lambda}
+\eta_{\nu 0}\eta_{\lambda 0}\partial_{\mu}\partial_{\rho}
+\eta_{\nu 0}\eta_{\rho 0}\partial_{\mu}\partial_{\lambda}\right.
\nonumber \\
&&\left.\makebox[8em]{}+\eta_{\mu 0}\eta_{\nu 0}\partial_{\lambda}
\partial_{\rho}+\eta_{\lambda 0}\eta_{\rho 0}\partial_{\mu}\partial_{\nu}
\right.
\nonumber \\
&&\left.+2\left(\eta_{\mu 0}\eta_{\nu 0}\eta_{\lambda 0}\partial_{0}
\partial_{\rho} 
+\eta_{\mu 0}\eta_{\nu 0}\eta_{\rho 0}\partial_{0}\partial_{\lambda}
+\eta_{\mu 0}\eta_{\lambda 0}\eta_{\rho 0}\partial_{0}\partial_{\nu}
+\eta_{\nu 0}\eta_{\lambda 0}\eta_{\rho 0}\partial_{0}\partial_{\mu}
\right)\right.
\nonumber \\
&&\left.\makebox[8em]{}+4\eta_{\mu 0}\eta_{\nu 0}\eta_{\lambda 0}
\eta_{\rho 0}\partial^{2}_{0}\right\}\tilde{D}^{(n)}(x-y)
\nonumber \\
&&+\frac{i}{2}\left(1-\frac{2}{n}\right)\eta_{\mu 0}\eta_{\nu 0}
\eta_{\lambda 0}\eta_{\rho 0}D^{(n)}(x-y)
\end{eqnarray}
where we used 
\[
\triangle\tilde{\tilde{D}}^{(n)}=\partial^{2}_{0}\tilde{\tilde{D}}^{(n)}
+\tilde{D}^{(n)}\;\;.
\]
Summing up all the terms we get 
\begin{eqnarray}
\lefteqn{\left[h_{\mu\nu}(x),\;h_{\lambda\rho}(y)\right]}
\nonumber \\
&&=\frac{i}{2}\left(\eta_{\mu\lambda}\eta_{\nu\rho}
+\eta_{\mu\rho}\eta_{\nu\lambda}
-\frac{2}{n}\eta_{\mu\nu}\eta_{\lambda\rho}\right)D^{(n)}(x-y)
\nonumber \\
&&+\frac{i}{4}\left\{-\eta_{\mu\lambda}\partial_{\nu}\partial_{\rho}
-\eta_{\mu\rho}\partial_{\nu}\partial_{\lambda}
-\eta_{\nu\lambda}\partial_{\mu}\partial_{\rho}
-\eta_{\nu\rho}\partial_{\mu}\partial_{\lambda}
+\frac{4}{n}\left(\eta_{\mu\nu}\partial_{\lambda}\partial_{\rho}
+\eta_{\lambda\rho}\partial_{\mu}\partial_{\nu}\right)\right\}
\tilde{D}^{(n)}(x-y)
\nonumber \\
&&+\frac{i}{2}\left(1-\frac{2}{n}\right)\partial_{\mu}\partial_{\nu}
\partial_{\lambda}\partial_{\rho}\tilde{\tilde{D}}^{(n)}(x-y)\;\;.
\end{eqnarray}

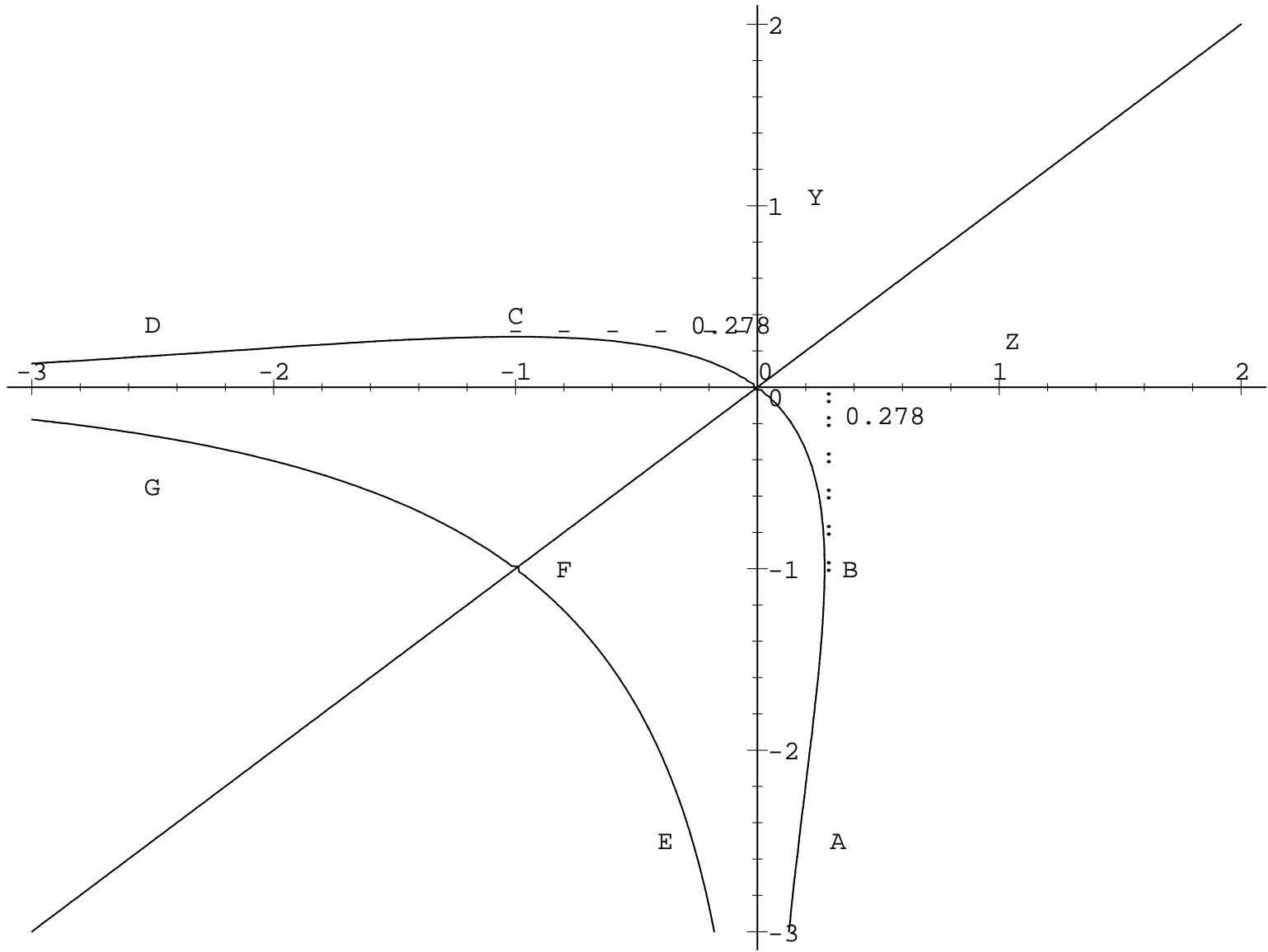
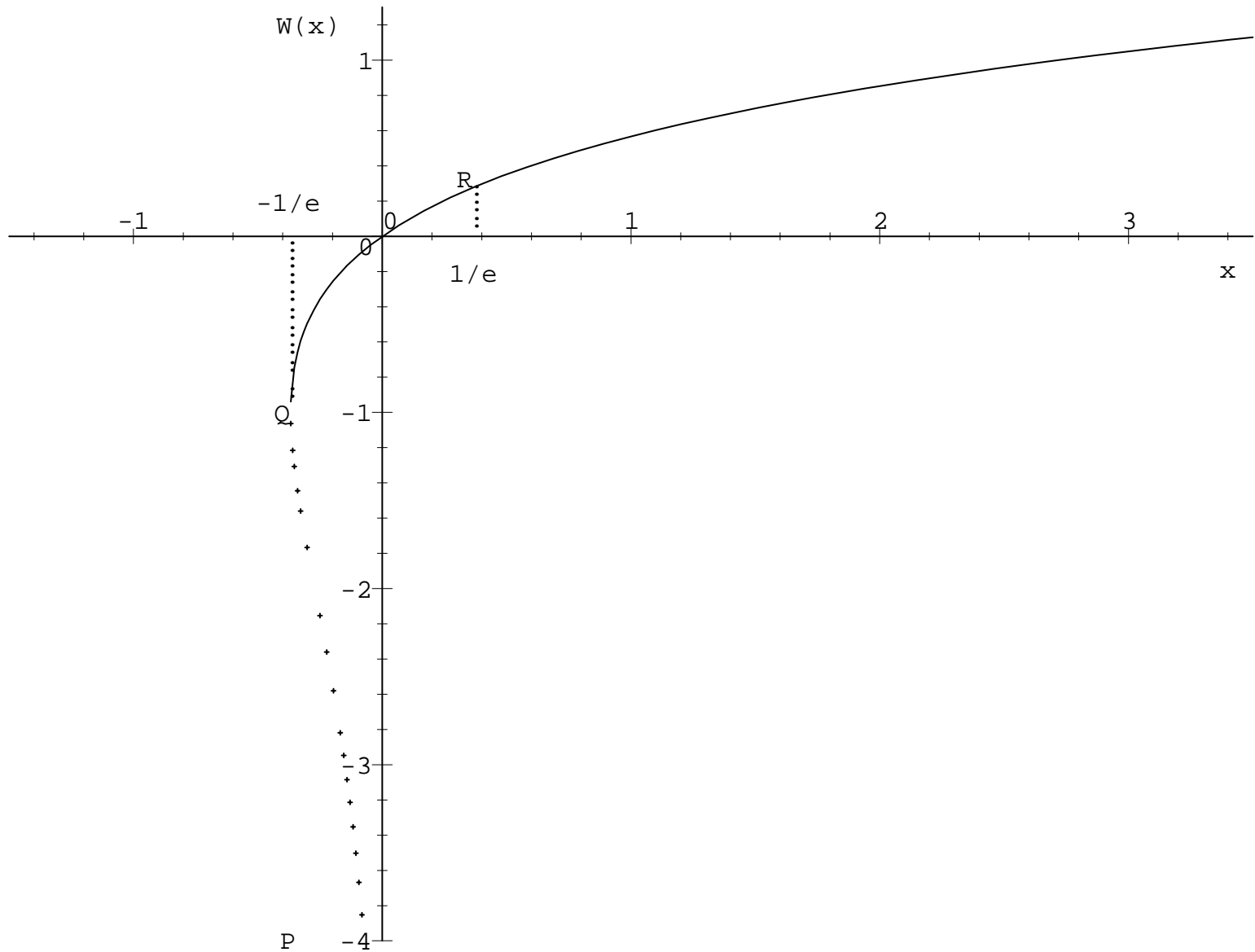
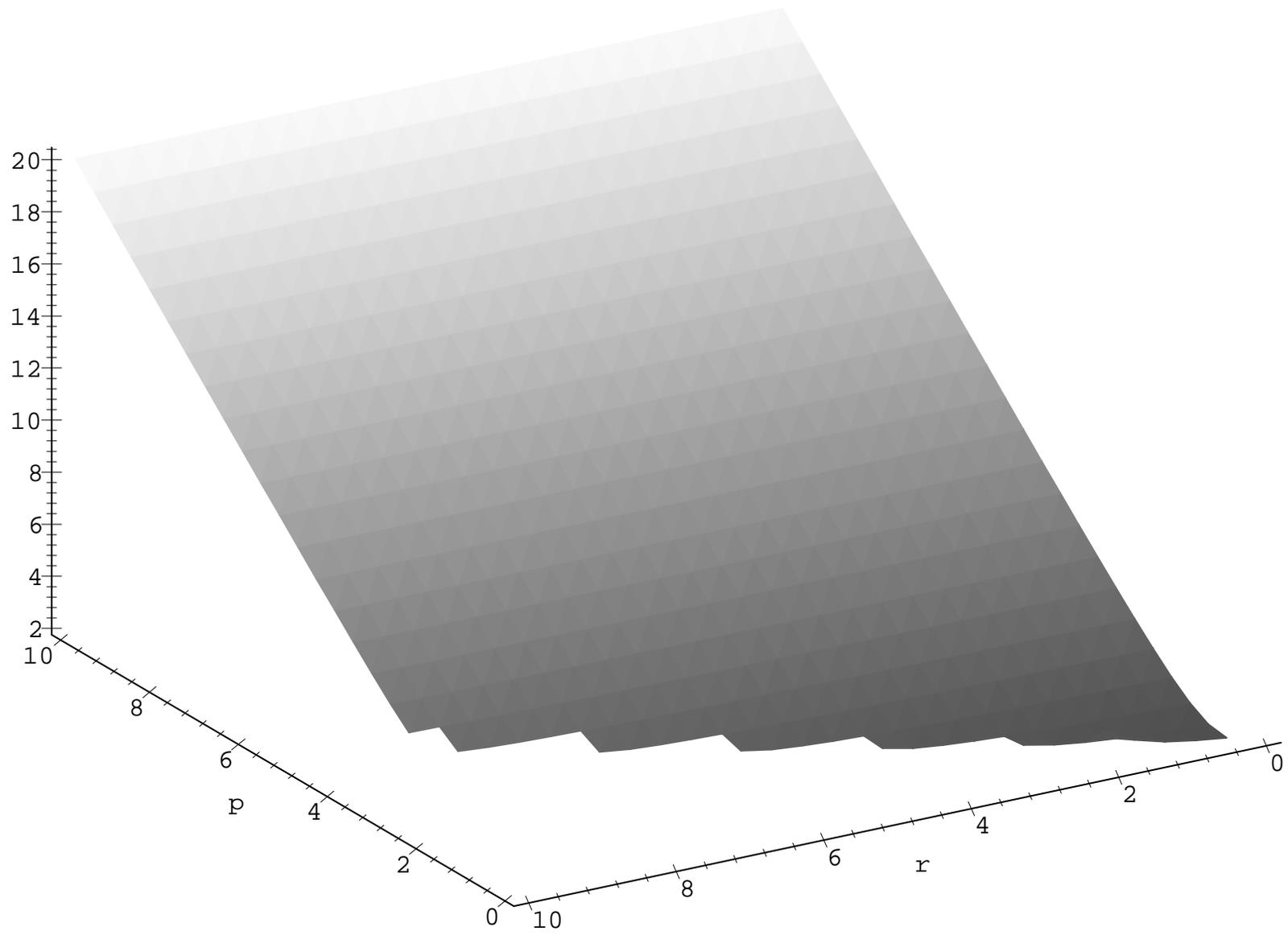
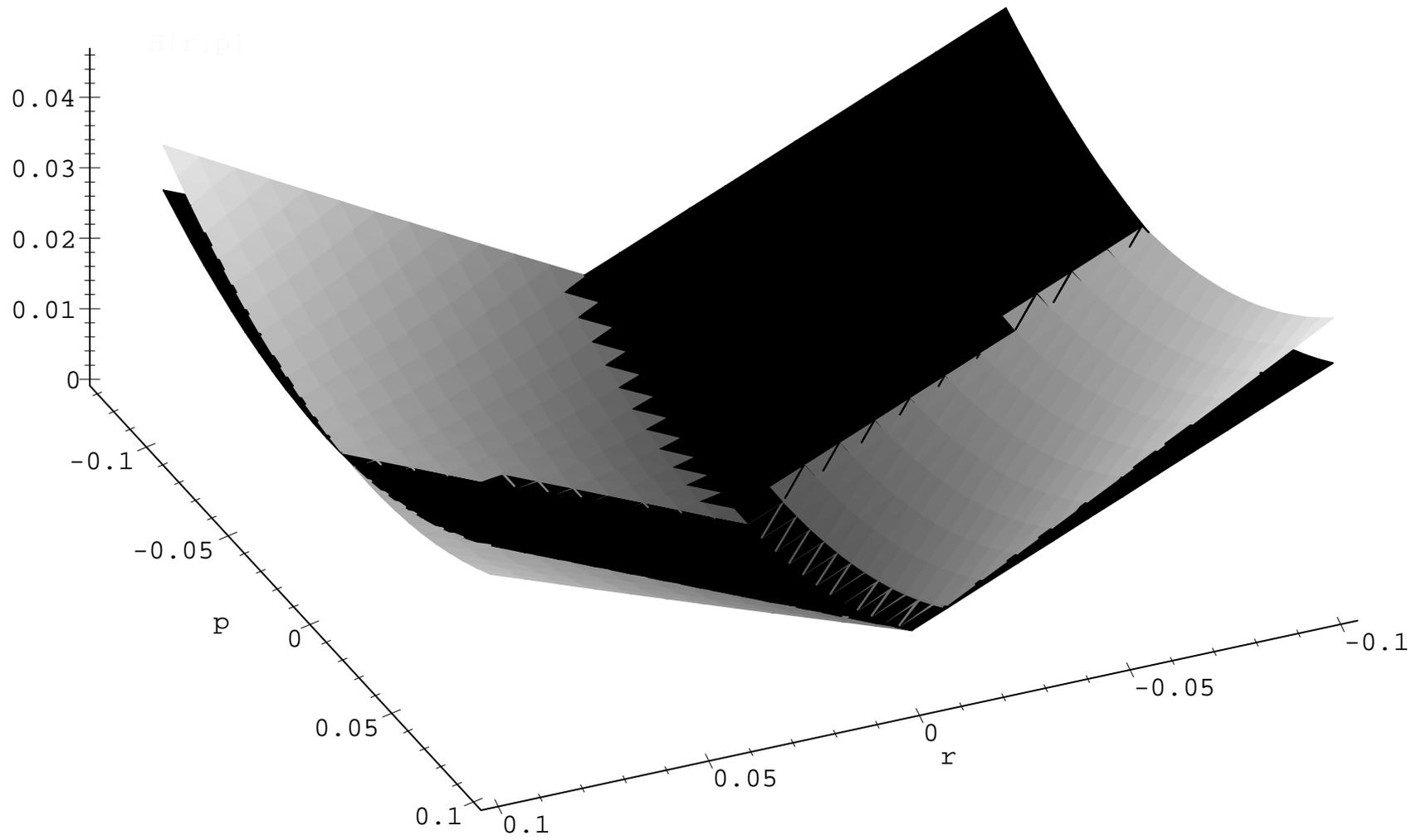
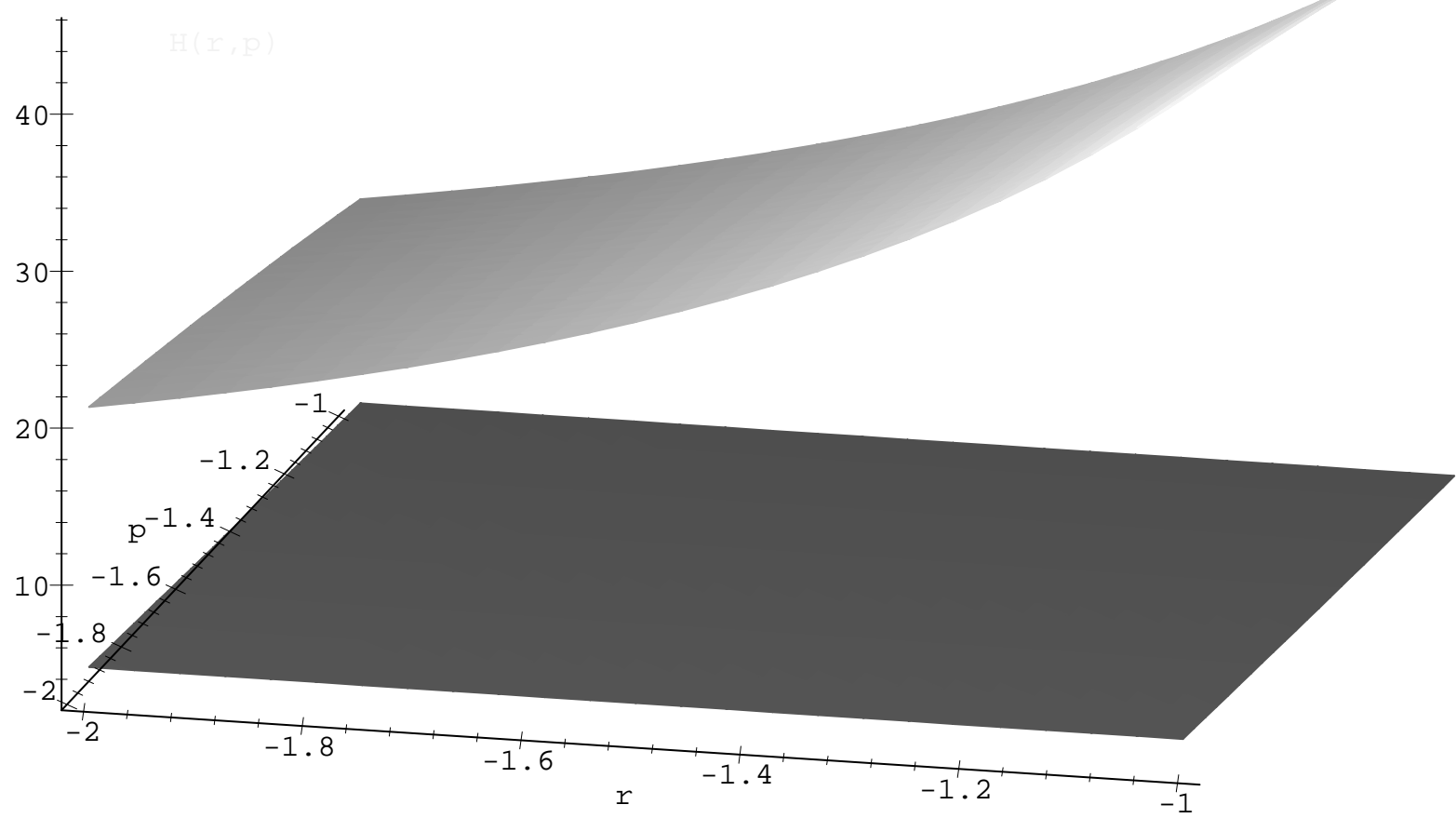
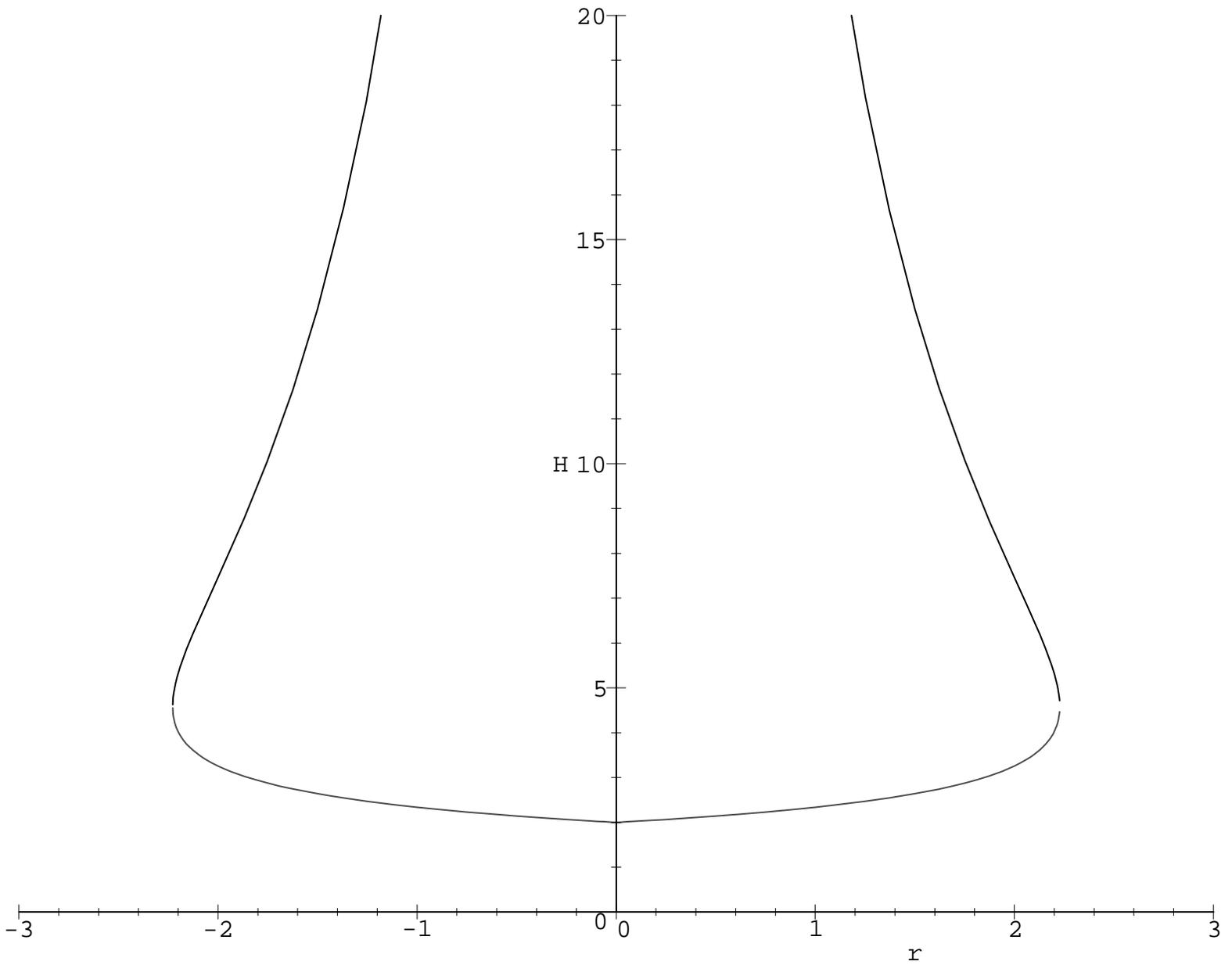
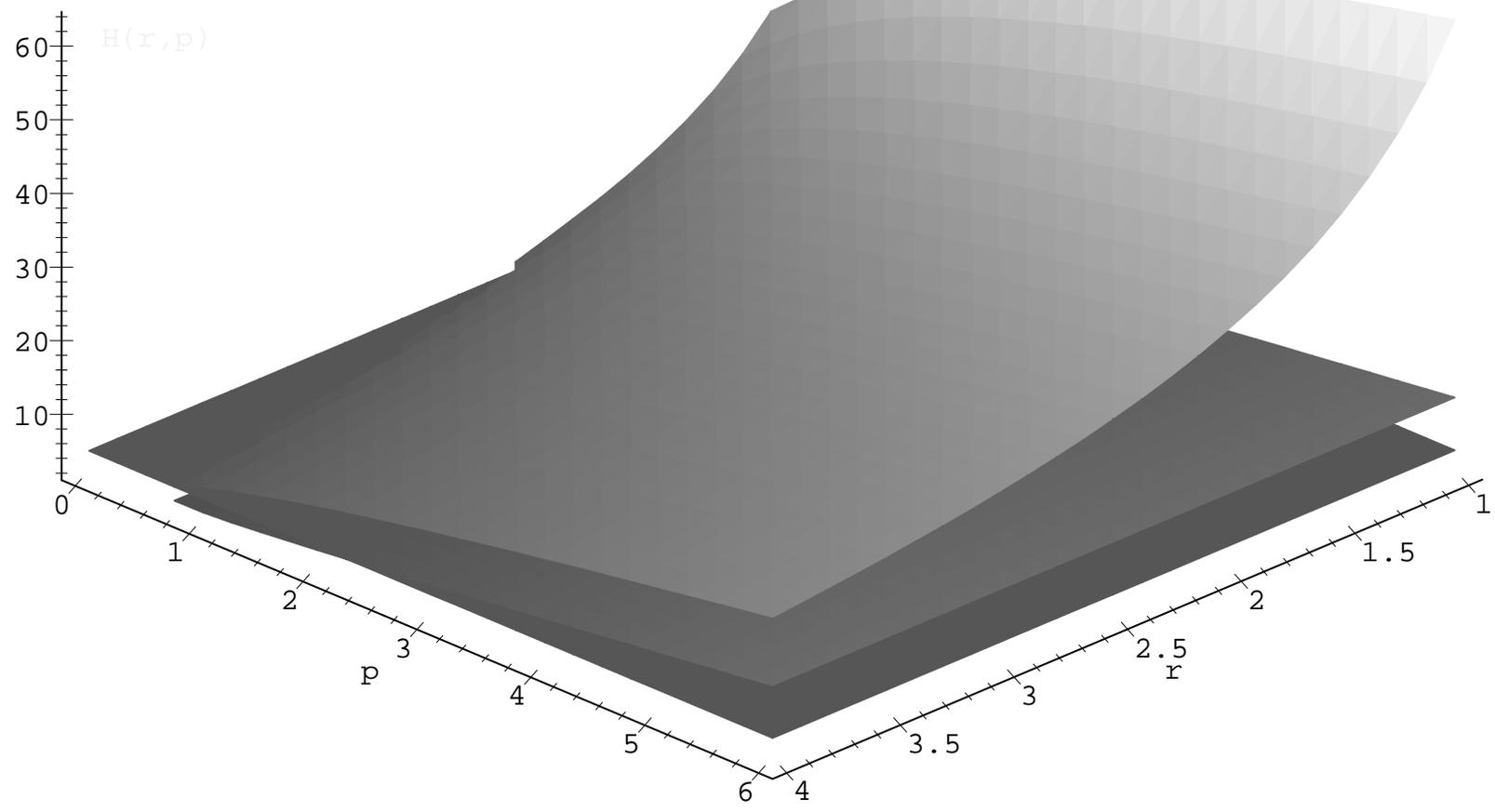
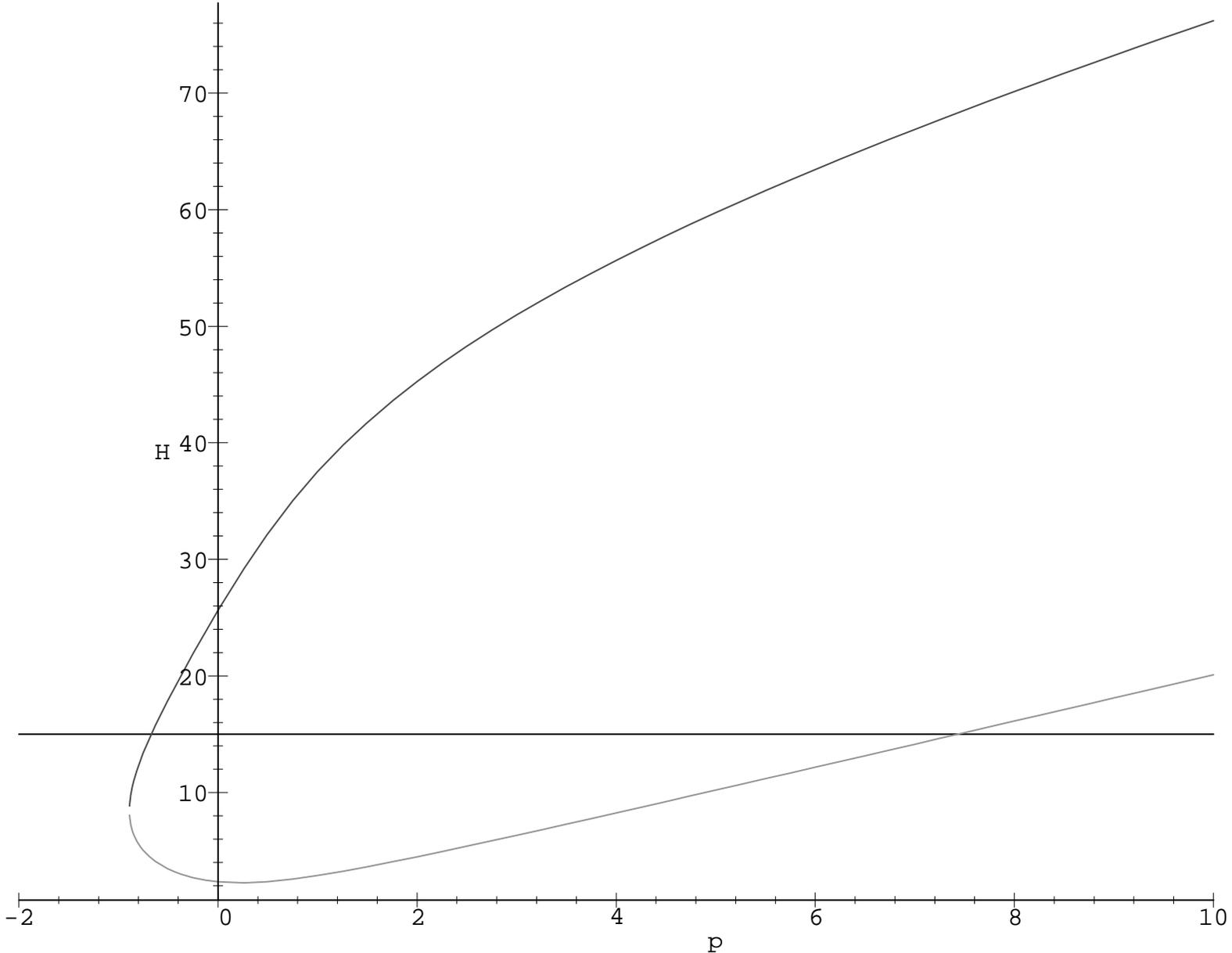
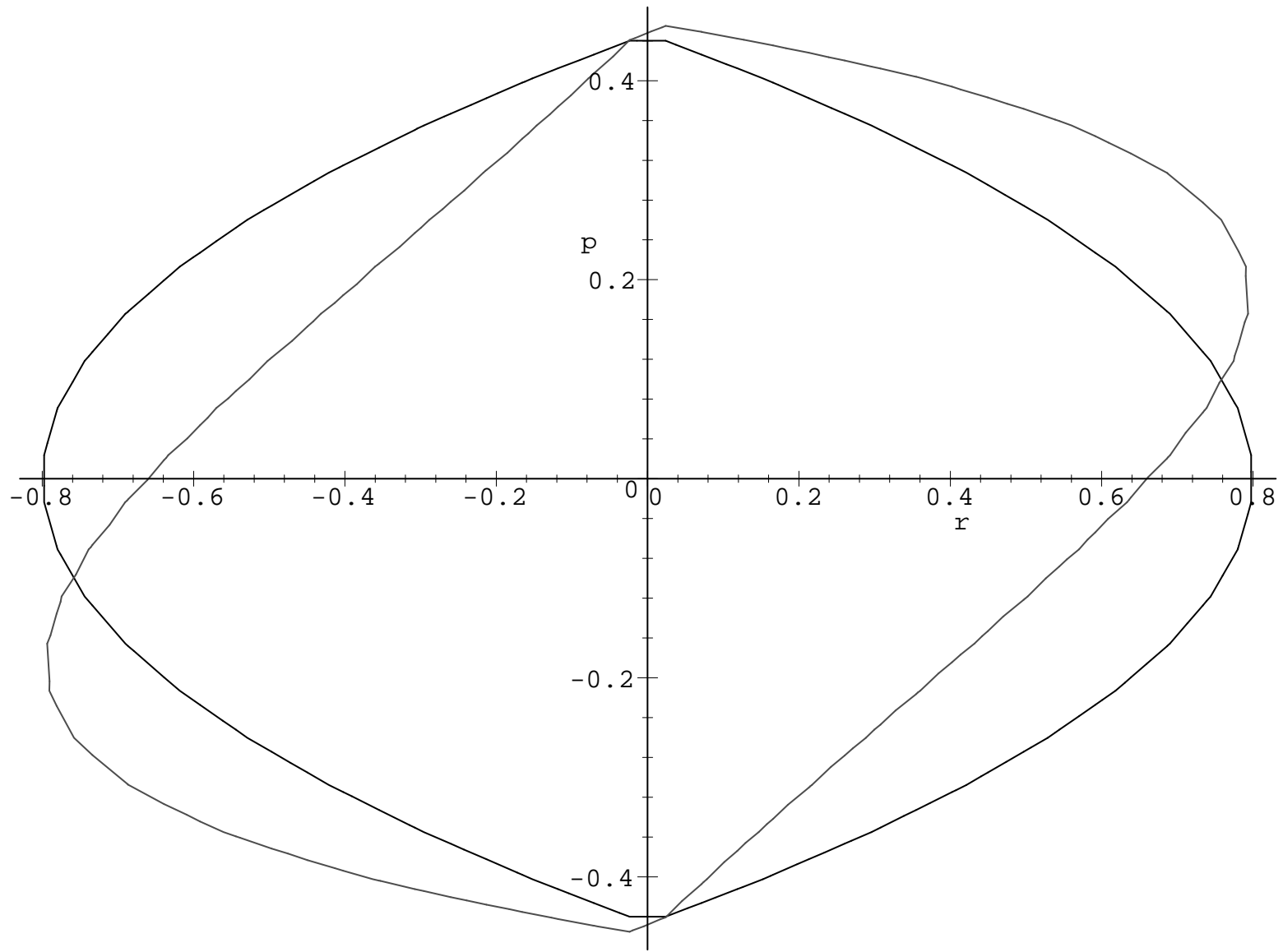
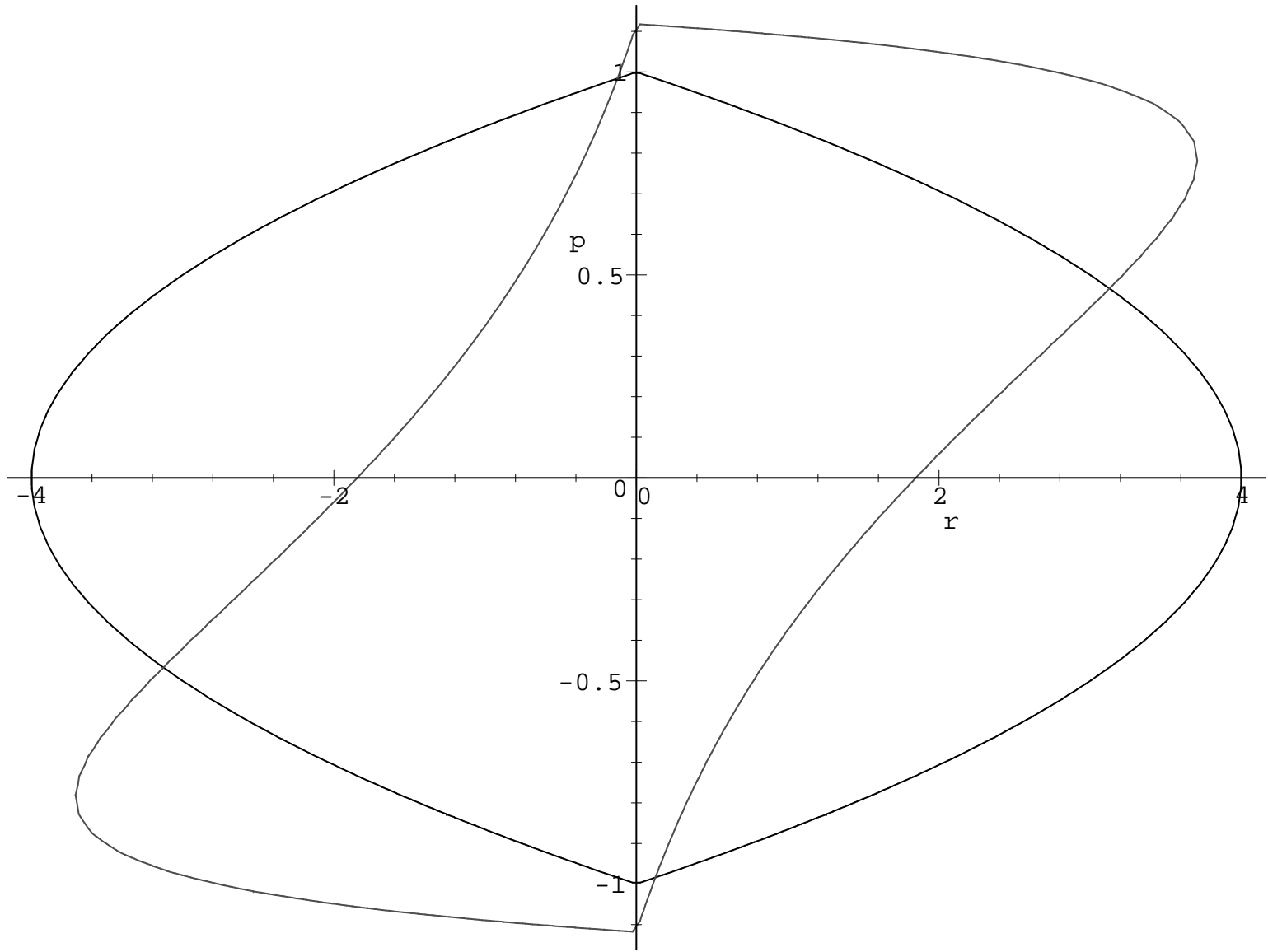
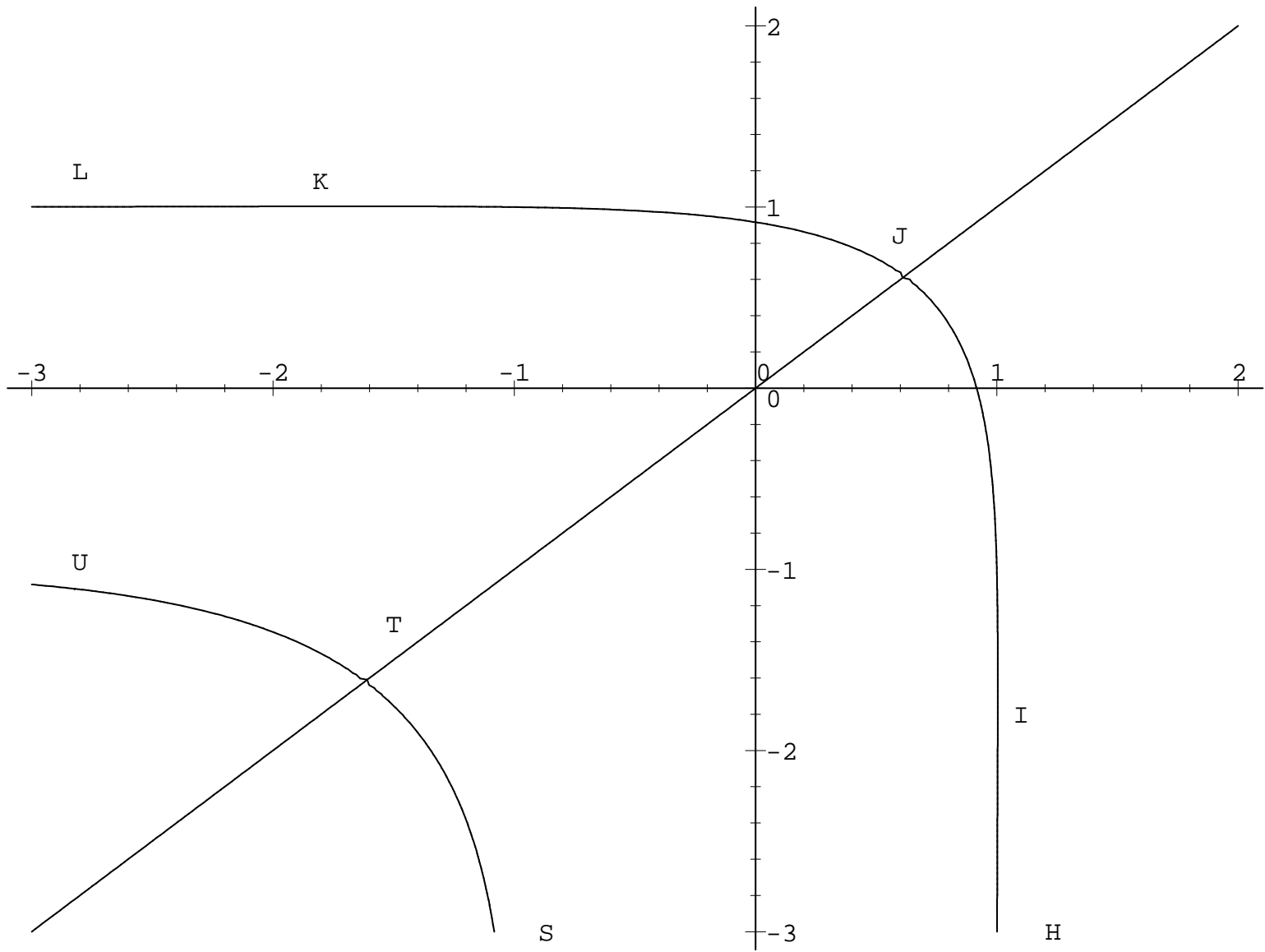
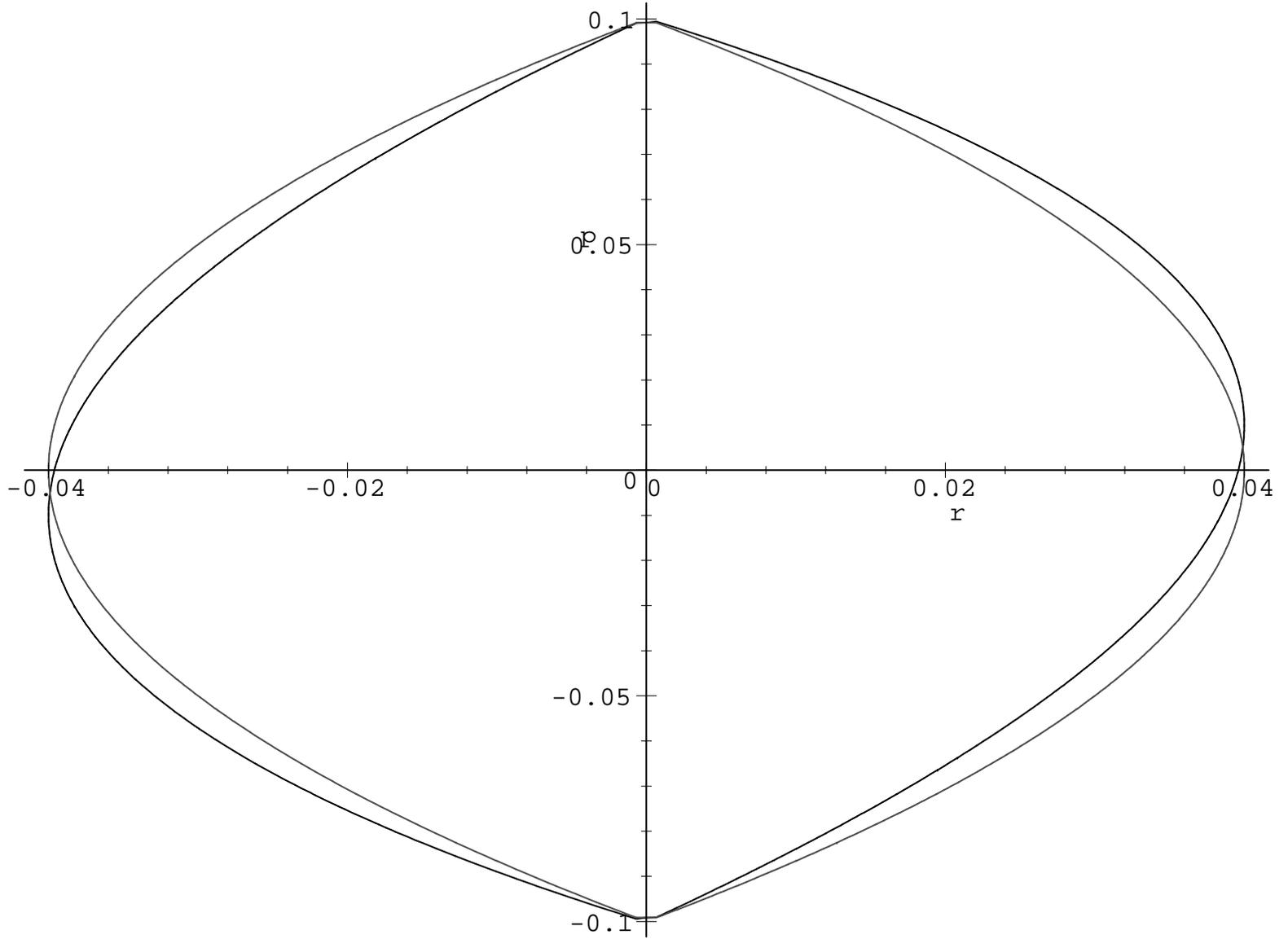
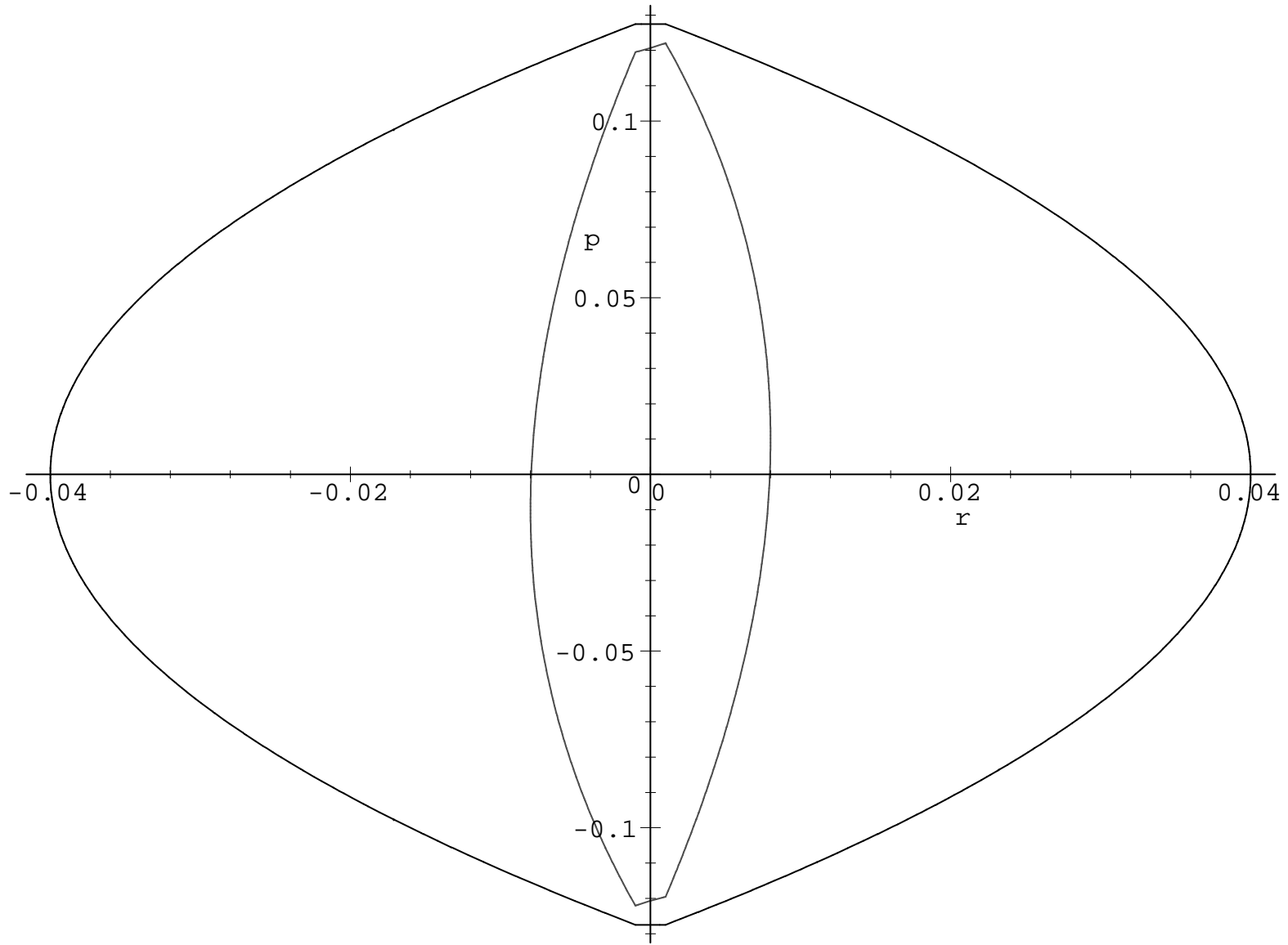
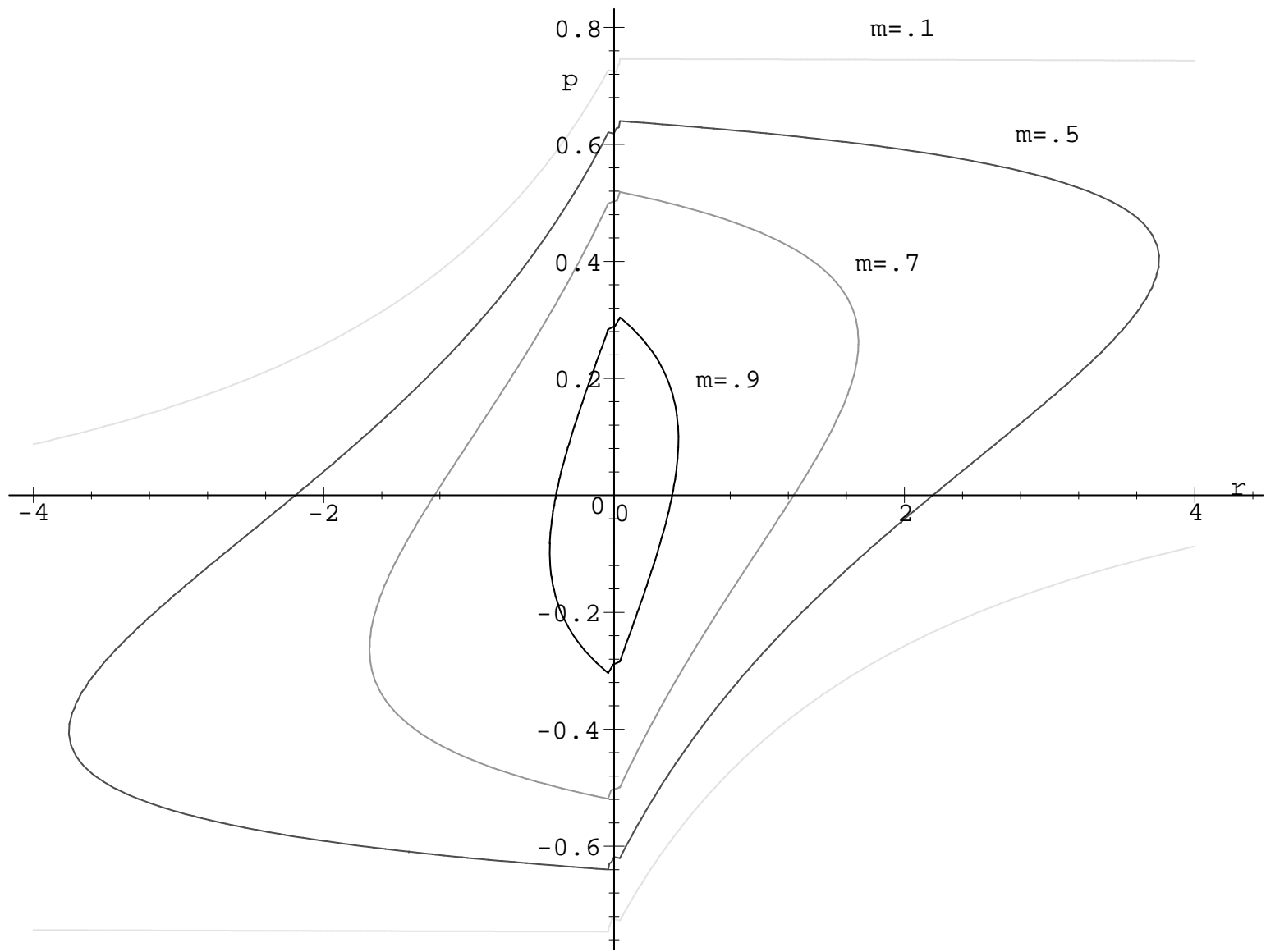
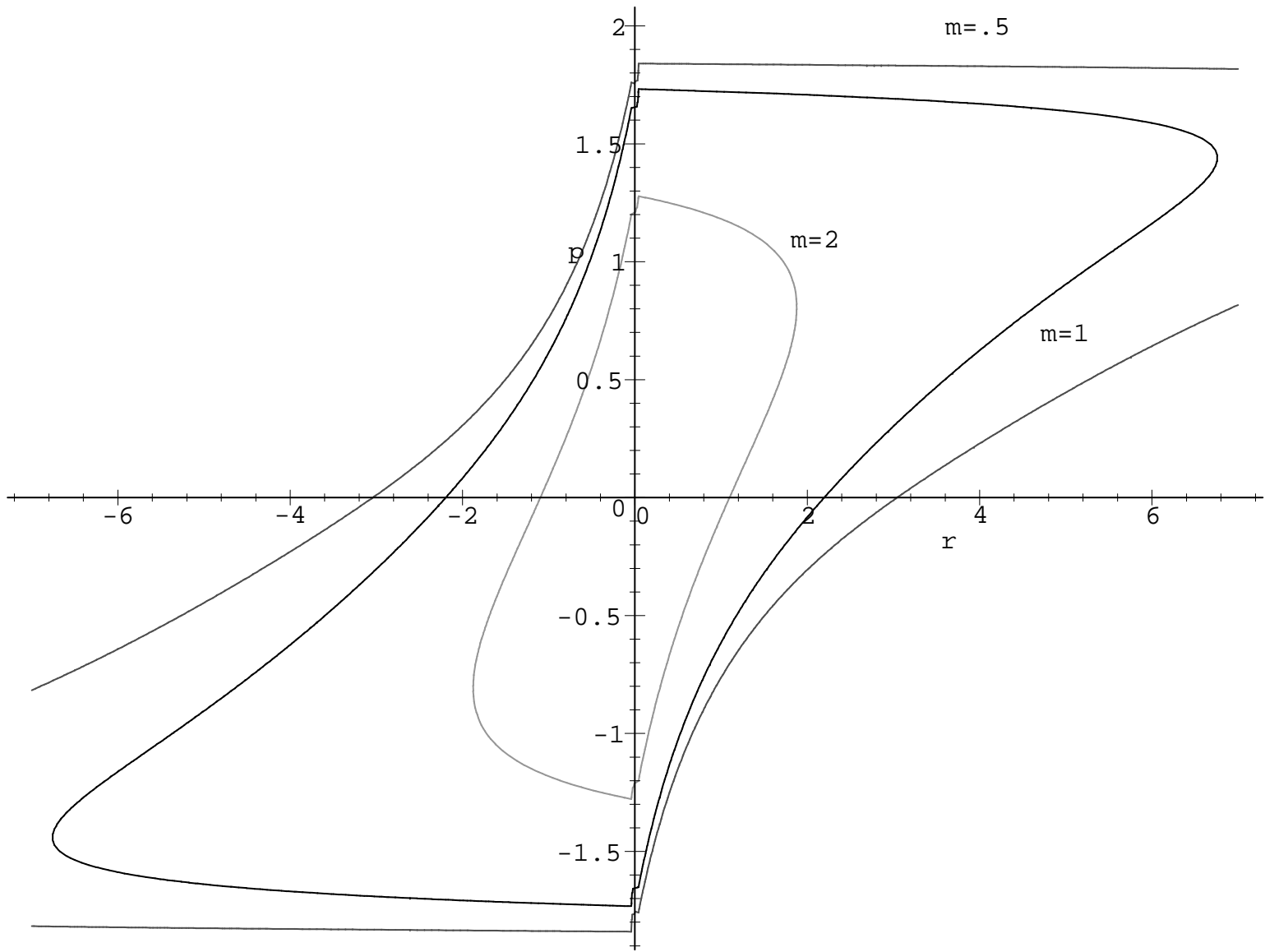
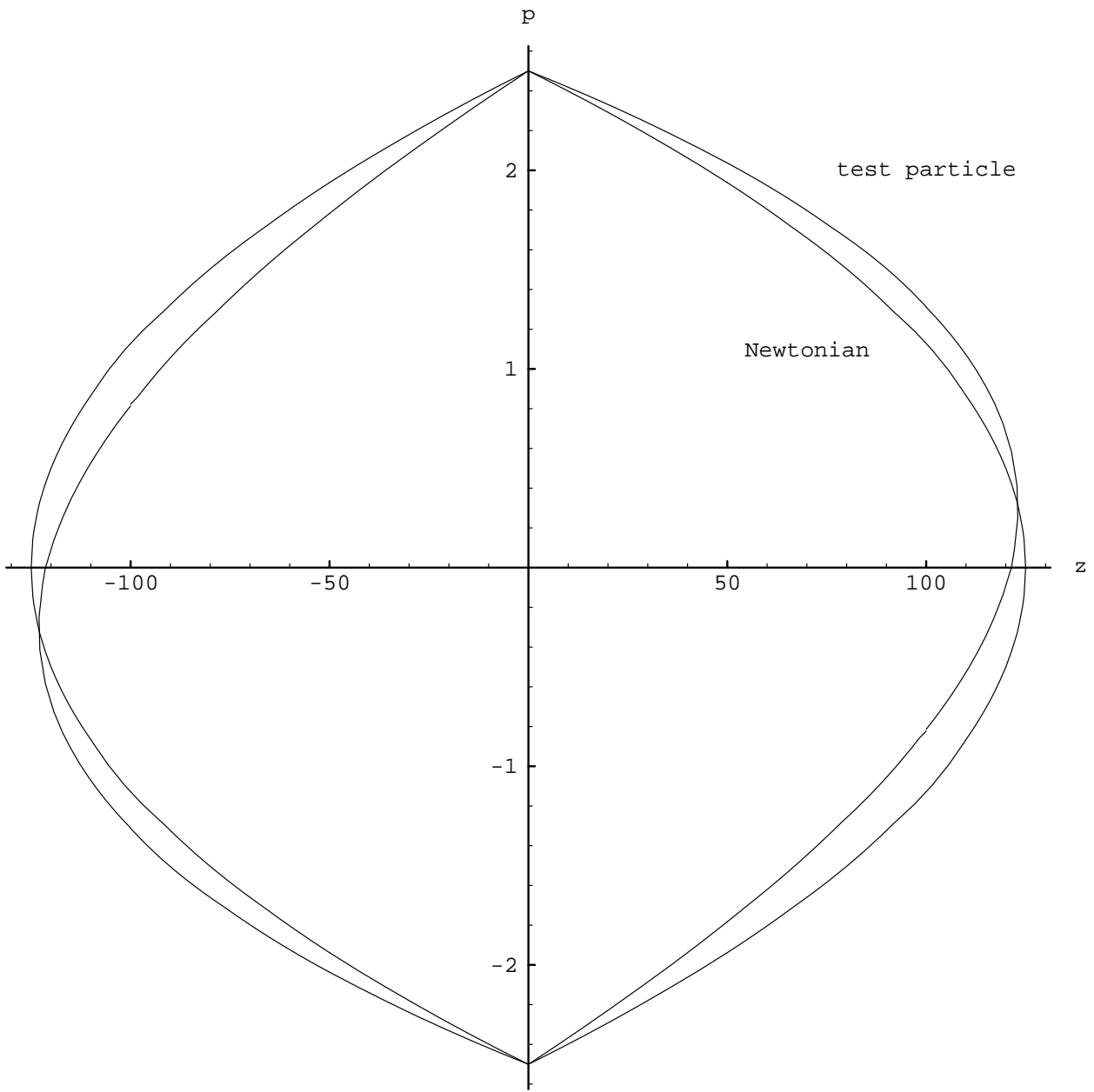
\end{document}